\documentclass[draft,journal,onecolumn,11pt,twoside]{IEEEtran}
\usepackage[dvips,final]{graphicx}
\usepackage{latexsym}
\usepackage{amssymb}
\usepackage[cmex10]{amsmath}
\usepackage{amsthm}
\usepackage{bm}
\usepackage{array}
\usepackage{balance}
\usepackage{algorithmic}
\usepackage{algorithm}
\usepackage[caption=false,font=footnotesize]{subfig}
\usepackage{multirow}

\normalsize

\theoremstyle{definition}
\newtheorem{example}{\em Example}

\theoremstyle{definition}
\newtheorem{defn}{\em Definition}

\theoremstyle{definition}

\theoremstyle{definition}

\title{Distributed Rate Allocation in Inter-Session Network Coding}

\author{Eirina~Bourtsoulatze,~\IEEEmembership{Student Member,~IEEE,}
	     Nikolaos~Thomos,~\IEEEmembership{Member,~IEEE,} 
              and~Pascal~Frossard,~\IEEEmembership{Senior Member,~IEEE} %
\thanks{E. Bourtsoulatze and P. Frossard are with the Signal Processing Laboratory 4 (LTS4), Ecole Polytechnique F\'{e}d\'{e}rale  de Lausanne (EPFL), Lausanne, Switzerland (e-mail: eirina.bourtsoulatze@epfl.ch; pascal.frossard@epfl.ch). N. Thomos is with the Signal Processing Laboratory 4 (LTS4), Ecole Polytechnique F\'{e}d\'{e}rale  de Lausanne (EPFL), Lausanne, Switzerland, and the Communication and Distributed Systems laboratory (CDS), University of Bern, Bern, Switzerland (e-mail: nikolaos.thomos@epfl.ch).}%
}
\begin{document}

\maketitle


\begin{abstract}
In this work, we propose a distributed rate allocation algorithm that minimizes the average decoding delay for multimedia clients in inter-session network coding systems. We consider a scenario where the users are organized in a mesh network and each user requests the content of one of the available sources. We propose a novel distributed algorithm where network users determine the coding operations and the packet rates to be requested from the parent nodes, such that the decoding delay is minimized for all the clients. A rate allocation problem is solved by every user, which seeks the rates that minimize the average decoding delay for its children and for itself. Since the optimization problem is a priori non-convex, we introduce the concept of equivalent packet flows, which permits to estimate the expected number of packets that every user needs to collect for decoding. We then decompose our original rate allocation problem into a set of convex subproblems, which are eventually combined to obtain an effective approximate solution to the delay minimization problem. The results demonstrate that the proposed scheme eliminates the bottlenecks and reduces the decoding delay experienced by users with limited bandwidth resources. We validate the performance of our distributed rate allocation algorithm in different video streaming scenarios using the NS-3 network simulator. We show that our system is able to take benefit of inter-session network coding for simultaneous delivery of video sessions in networks with path diversity. 
\end{abstract}

\vspace{-0.2cm}


\section{Introduction}
\label{sec:intro}

Distributed network architectures and protocols have gained much popularity over the past few years due to their scalability properties. Deployed initially for file sharing, today distributed systems are exploited for more demanding network applications such as live streaming, VoD, multi-party conferencing {\em etc}. The essential advantage of these systems over the traditional client-server architecture is their ability to sustain a large number of users without increasing the server load, as users contribute their upload bandwidth to the system. This, however, comes at the cost of dynamic and unpredictable behavior of the network nodes. It renders the centralized routing methods challenging and necessitates distributed algorithms for data delivery. In this context, network coding \cite{Ahlswede00} has been considered recently as a solution to improve the performance of distributed systems. It removes the need for content reconciliation among the users and offers decentralized control as well as efficient adaptation to bandwidth variations and losses. 

A broad spectrum of distributed algorithms that utilize network coding has been proposed in the literature. These works mainly focus on the case where a single data source from one or more servers is delivered to multiple users. It is common, however, that the network resources need to be shared by concurrent applications. In such settings, inter-session network coding \cite{HoNC} arises as a natural extension of network coding for efficient use of network resources with multiple sessions. Yet, the design of the network codes is not a trivial task; random mixing of all the sessions that exist in the network may lead to significant increase in the decoding delay for users that recover their source of interest from the combinations of different sessions. 

In this paper, we build on our previous work \cite{BourtsouPV2012} and address the problem of designing a distributed rate allocation algorithm that decides how many packets of each session combination should be transmitted on the network links. We consider a scenario with concurrent sessions that transmit data to users organized in a mesh network. The proposed protocol is receiver-driven and comprises two steps. First, the node requests and receives information about its local neighborhood that is formed of parents and children nodes. Second, the node requests intra- and inter-session network coded packets at specific rates. These rates are obtained by solving an optimization problem that seeks for the optimal rate allocation among different packet combinations. The objective of the optimization algorithm solved at each node is to minimize the average decoding delay of the node and its children nodes.

The delay minimization problem is a priori non-convex. We approximate it with a set of convex subproblems by introducing the new concept of equivalent flows. An equivalent flow is defined for every component session of an inter-session combination. It can be regarded as a hypothetical flow with a rate equal to the innovative rate for the component session. This leads to an estimation of the expected number of packets necessary for decoding a source of interest from a particular packet combination. Based on the equivalent flows representation, the original optimization problem is decomposed into several convex rate allocation subproblems that are easily solvable. Their solutions are then combined to yield an approximate yet effective solution to the optimal rate allocation. Simulation results demonstrate that the proposed scheme eliminates the bottlenecks and reduces the decoding delay of users with limited resources, while it enables the timely delivery of time sensitive data. The benefits of our algorithm are finally validated by NS-3 simulations for video streaming in different network scenarios.  

In summary, the main contributions in this paper are the following:
\begin{itemize}
\item we propose a new formulation of a decoding delay optimization problem for inter-session network coding in wired overlay networks,
\item we introduce the novel concept of equivalent flows for approximate delay computation in inter-session network coding scenarios,
\item we design a new distributed rate allocation algorithm for minimizing the decoding delay. We validate the performance of our algorithm in video streaming scenarios with help of a network simulator.
\end{itemize}

The rest of the paper is organized as follows. In Section \ref{sec:relatedwork} we discuss the related work. We describe the scenario that we consider and the communication protocol in Section \ref{sec:systemoverview}. In the same section, we formulate the distributed rate allocation problem with inter-session network coding. The concept of equivalent flows is introduced in Section \ref{sec:equivalentflows}. Our proposed distributed algorithm for delay minimal rate allocation is presented in Section \ref{sec:ratealloc}. In Sections \ref{sec:evaluation}, we evaluate the performance of the proposed rate allocation scheme in terms of the average decoding delay, while in Section \ref{sec:video} we present the results of the video streaming simulations. Section \ref{sec:conclusions} concludes our work.


\section{Related work}
\label{sec:relatedwork}

With the recent advances in network coding research, the potentials of network coding have developed in the framework of P2P and overlay data delivery networks \cite{FrossardNCOverview}. For example, Wang {\em et al.} \cite{WangR207} have proposed a design called {\em R\textsuperscript 2} that combines the random push strategy with random network coding. The work in \cite{MedardQoE} provides an analysis of the rate-delay-reliability trade-offs in a P2P streaming system. The authors derive upper and lower bounds on the minimum initial buffering required so that the playback interruption probability remains below a certain level. Network coding has also been considered for unequal error protection in overlay streaming systems as in \cite{ ThomosUEPTMM2011}, where the authors propose a distributed receiver-driven algorithm for prioritized media delivery.

Apart from the single session streaming cases, network coding has been also considered for multiple concurrent unicast and multicast scenarios. It has been shown that linear network coding is not sufficient for achieving the capacity bound \cite{Dougherty05}; however, significant throughput gains can still be obtained with linear inter-session network coding as shown in \cite{COPE}, which describes an implementation of opportunistic network coding for multiple unicast flows over wireless networks. Recently, several inter-session network coding algorithms have been proposed, mainly for data delivery in wireless networks \cite{SefDelayOptNC,KhreishahJSAC09, SefI2NC,FlowBasedXORLossy, VirtualMulticasts}. Some of the works extend the COPE architecture \cite{COPE} by considering application-specific features when designing the network codes. The work in \cite{SefDelayOptNC} for example studies the benefits of delaying packets at intermediate nodes in order to create more network coding opportunities. The proposed network coding scheme builds on COPE and incorporates an optimization framework that seeks for the optimal code and transmission policies that optimize the rate-distortion function. The performance of COPE and COPE-based systems degrades significantly in the presence of losses and network coding is turned off when the packet loss rate reaches a certain threshold. To deal efficiently with the packet losses, the authors in \cite{SefI2NC} propose a joint application of intra-session and inter-session network coding. Intra-session network coding is used for protection against packet losses, whereas inter-session network coding increases the throughput of the network. In order to characterize the capacity achieved with inter-session network coding for the 2-hop relay networks in the presence of losses, a flow based analysis is presented in \cite{FlowBasedXORLossy}. The key idea is to regard packets as members of flows and not as independent entities as in \cite{COPE, SefDelayOptNC,SefI2NC}. A different approach for finding the feasible rate region is built on virtual multicasts \cite{VirtualMulticasts}. The flow-based problem formulation stated in \cite{VirtualMulticasts} provides a rate region which is at least as large as the rate region that can be achieved without inter-session network coding. 

While the benefits of inter-session network coding are well understood in the wireless scenarios, in wireline networks the construction of practical inter-session network coding algorithms is more challenging. The reason lies in the difference between the two communication media. The broadcast nature of wireless channels promotes the application of inter-session network coding through overhearing \cite{COPE,SefDelayOptNC}, {\em i.e.}, packets that are required for decoding can be overheard without wasting additional resources and decoding can be performed at every hop. This is not the case in wireline networks. Various theoretical aspects of inter-session network coding, such as sufficiency of linear codes and complexity of identifying coding opportunities, are studied in \cite{pairwiseISNCShroff2010} for the special case of pairwise coding in wireline networks. Kim {\em et al.} \cite{EvolutionaryISNC} propose a more generic solution that utilizes linear network coding and does not restrict the codes to specific classes such as pairwise or XOR coding. The coding strategy is determined with the help of Genetic Algorithms that optimize a certain cost objective. The work in \cite{LunTIT2011} provides a different perspective on the design of inter-session network coding algorithms by exploiting the queue-length information to make the scheduling-routing-coding decisions. In \cite{SaltarinICME2011}, the authors propose a low-complexity receiver-driven P2P system for delivery of multiple description coded data, that combines Raptor codes with intra- and inter-session network coding. 

To the best of our knowledge, there is however no work in the literature that addresses the problem of minimizing the average decoding delay in wireline mesh networks by distributed rate allocation in inter-session network coding.


\section{Data delivery with inter-session network coding}
\label{sec:systemoverview}


\subsection{Framework}
\label{sec:setup}

\begin{figure}[t]
	\begin{center}
		~\includegraphics[width=0.35\textwidth]{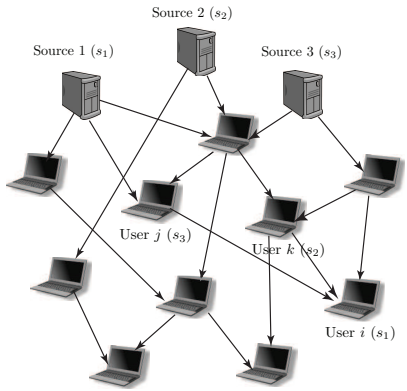}~
	\end{center}
	\caption{Illustration of a multi-session scenario. Each source provides different data to the network. The users are organized in a mesh network, where each user requests a specific source data.\label{fig:p2ptopology}}	
\end{figure}

We consider a set of sources $\mathcal{S}$ and a set of users $\mathcal{N}$  that request data from different sources. The source data is segmented into blocks of $N_{s}$ packets, and the sources transmit simultaneously at rate $U_{s}$, $s\in\mathcal{S}$. The users are organized in a wireline mesh network. The network is assumed to be directed and free of cycles. It is modeled as a directed acyclic graph $\mathcal{G} = (\mathcal{V},\mathcal{E})$, where $\mathcal{V} = \mathcal{S}\cup\mathcal{N}$ represents the set of network nodes, and $\mathcal{E}$ is the set of connecting links between the network nodes. The directed link connecting any two nodes $i$ and $j$ is denoted as $(i,j) \in \mathcal{E}$. It is characterized by the link capacity $b_{ij}$ expressed in packets/sec and the average packet loss probability $\pi_{ij}$. If nodes $i$ and $j$ are connected with the directed link $(i,j)$, we call node $j$ as a child of node $i$, and node $i$ is called the parent of node $j$. An example of such mesh network is illustrated in Fig.~\ref{fig:p2ptopology}.

The network nodes represent end users, who are interested in receiving only one of the source data, but also act as relay nodes. Since the upload bandwidth of the sources is limited and only a small number of users can acquire the requested packets directly from the sources, the majority of the network users are served by their parent nodes. This implies that a user may request and forward not only packets of the source that it has subscribed to, but also packets that are useful for its children nodes. In order to increase the network throughput and alleviate the bottlenecks created by the limited network resources, we propose to allow the network nodes to implement inter-session network coding. Inter-session network coding \cite{HoNC} is an extension of network coding \cite{Ahlswede00} to the case of multiple concurrent sessions (data sources) that share the same network resources. It essentially consists in combining packets from different sessions (sources), contrarily to intra-session network coding where only packets of the same session (source) participate in the packet combinations. When linear operations are considered, an inter-session network coded packet can be formally represented as 
\begin{equation}
	y  = \sum \limits_{s = 1} ^ {|\mathcal{S}|} 
	\sum \limits_{l = 1} ^ {N_{s}} a_{s,l}x_{s,l}
\label{eq:A}
\end{equation}
where all the operations are performed in a Galois field of size $q$, GF$(q)$. The $l$-th original source packet of the $s$-th session is denoted as $x_{s,l}$ and $a_{s,l}$ is the corresponding coding coefficient. It should be noted that not all of the sessions necessarily participate in a particular inter-session network coded packet. When some of the sessions are not included in the combination, the corresponding coding coefficients are zero. From that perspective, intra-session network coded packets can be viewed as a special case of inter-session network coding where packets from only one session participate in the coding operations.

Depending on the available set of packets at the parent nodes, every user may request intra-session network coded packets of its session of interest, as well as inter-session network coded packets, {\em i.e.}, packets that are combinations of different sessions. These combinations do not necessarily involve packets from the session requested by the user. 

We denote as $\mathcal{T}$ the set of all the possible packet types that can be generated in the network. Every element $t$ of $\mathcal{T}$ represents a particular combination of sessions. Hence, in a network with $|\mathcal{S}|$ concurrent sessions, the number of different packet types is $2^{|\mathcal{S}|} -1$. Intra-session network coded packets are also included in the set $\mathcal{T}$. We denote as $\mathcal{T}_t$ the set of packet types that can be combined to generate coded packets of type $t$. The sessions that participate in a particular combination of packets $t$ form the set $\mathcal{S}_t$. We will refer to the sessions in the set $\mathcal{S}_t$ as the \textit{component sessions} of flow of type $t$. We also define the sets $\mathcal{T}^{s}$ and $\mathcal{T}_{t,s}$. The set $\mathcal{T}^{s}$ is a subset of $\mathcal{T}$ and contains the packet types that have session $s$ as a component session, {\em i.e.}, $\mathcal{T}^{s} = \{t\in\mathcal{T} : s\cap\mathcal{S}_{t} \neq \emptyset \}$.  The set $\mathcal{T}_{t,s}$ includes all the packet types $t^\prime \in \mathcal{T}_{t}$ that can be used to generate packets of type $t$ and have $s$ as a component session, {\em i.e.}, $\mathcal{T}_{t,s} = \{t^{\prime}\in\mathcal{T}_{t} : s\cap\mathcal{S}_{t^{\prime}} \neq \emptyset \}$. 

Every user, upon receiving a sufficient number of network coded packets, decodes the received packets in order to obtain the packets of the requested session. The decoding of a particular session is typically performed by means of Gaussian elimination when a full rank system of packets is received. Note that, since the local coding coefficients are drawn randomly according to a uniform distribution from the GF$(q)$, a header of length $\sum _{s\in \mathcal{S}}N_{s}\log(q)$ bits is appended to the network coded packets. This header identifies all the coding operations performed on the packets while they travel through the network; it renders the decoding process feasible, since the encoding structure becomes implicit.

In general, the application of inter-session network coding is not trivial. Random mixing of all the available sessions is not always efficient, as it may cause an unacceptable increase of the decoding delay for a specific source data. This is due to the fact that users need to receive enough innovative packets in order to decode all the encoded sessions along with the session of their interest. The term ``innovative'' refers to packets that bring novel information with respect to the packets that have been previously received by the node. These packets are linearly independent from the packets that are already stored in the node's buffer. In order to alleviate the shortcomings of the random mixing of all the sessions, an efficient rate allocation algorithm is essential. The goal of such rate allocation algorithm is to determine the sessions that should be combined and the rate that should be allocated to each combination in order to minimize the average decoding delay. This decoding delay depends on the innovative packet rates that the user receives for each of the session combinations that are available in the network. Since the networks are typically characterized by dynamics such as bandwidth variations, varying channel conditions, users' arrivals/departures at random time instances, {\em etc.}, a centralized rate allocation strategy is impractical.  Therefore, we propose to optimize the decoding delay locally in a small neighborhood that comprises the node itself and its parent and children nodes. The rate optimization is performed with only a partial knowledge of the network statistics and the required communication overhead is small. Due to the distributed nature of the problem, global optimality can however not be guaranteed anymore, but the solution proposed in this paper proves to be effective and adapted to realistic settings.


\subsection{Communication protocol}
\label{sec:protocol}

\begin{figure}[t]
	\begin{center}
		~\includegraphics[width=0.6\textwidth]{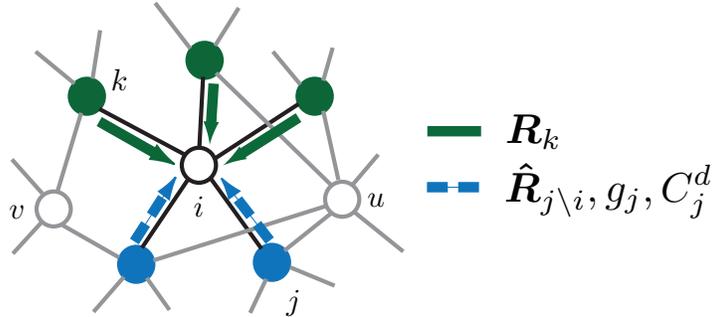}~
	\end{center}
	\caption{ Communication protocol. The neighborhood of the user consists of the parent nodes (in green) and the children nodes (in blue). The green arrows (solid) indicate the information communicated to the node $i$ by its parents. The blue arrows (dashed) represent the information received by the node from its children.\label{fig:protocol}}	
\end{figure}

The distributed delay optimization solution requires some exchange of information between the network users. We propose the following communication protocol. Let us consider the node $i$ and its local neighborhood that consists of the set of parent nodes $\mathcal{A}_i$ and the set of children nodes $\mathcal{D}_i$ as depicted in Fig.~\ref{fig:protocol}. We assume that the node $i$ is aware of the local network statistics, {\em i.e.}, the channel capacity and the loss rates  of the input and output links ($b_{ki}$, $\pi_{ki}$, $\forall k\in\mathcal{A}_i$ and $b_{ij}$, $\pi_{ij}$, $\forall j\in\mathcal{D}_i$, respectively). We also assume that every child node $j$ communicates to the node $i$ the identity $g_j$ of the session it wants to receive and its total input capacity $C_{j}^{d} = \sum _{u\in\mathcal{A}_{j}}b_{uj}$.

Whenever the user $i$ wants to optimize the requested packet flow rates, it requests the users in its neighborhood to provide all the necessary information about the local status of the system. Specifically, every parent $k\in\mathcal{A}_i$ sends to the node $i$ a vector $\bm{R}_k$ with the values of the input innovative flow rates for every packet type $t\in\mathcal{T}$.  Every element $R_k^t$ of this vector represents the total input innovative flow rate of packets of type $t$ available at the parent node $k$ at the time instant when the node $i$ performs the optimization of the rate allocation. In more details, $R_k^t$ is given as 
\begin{equation}
	R_k^t = \sum\limits_{n \in \mathcal{A}_k} r_{nk}^t, \quad \forall t \in \mathcal{T}
\label{eq:C}
\end{equation}
where $r_{nk}^t$ is the innovative rate of packets of type $t$ received by node $k$ from its parent node $n$.\footnote {Here we have assumed that two packets that arrive from two different links are innovative with respect to each other with high probability. This holds in general in networks with high path diversity, which is the case considered in this work.} Similarly, every child node $j \in \mathcal{D}_i$ forwards to the node $i$ a vector $\bm{\hat{R}}_{j\backslash i}$, with $\hat{R}_{j\backslash i}^t$ representing the total innovative input flow rate of packets of type $t$ that the node $j$ receives from its parents, except for the parent $i$
\begin{equation}
	\hat{R}_{j\backslash i}^t = \sum\limits_{u \in \mathcal{A}_j\backslash i} r_{uj}^t, \quad \forall t \in \mathcal{T}
\label{eq:E}
\end{equation}
The communication protocol is illustrated in Fig.~\ref{fig:protocol}.


\subsection{Distributed delay minimization problem}
\label{sec:problem}

We are now able to formulate the distributed rate allocation problem that is solved independently in every network node. It consists in determining the optimal innovative rates that the user requests from its parents so that the average expected delay of the user and its children is minimized. The reason for considering the children nodes in the rate allocation optimization performed by every network node is to avoid selfish behaviors of the users. It is obvious that if the node performed the rate allocation taking into account only its own delay, it would preferably allocate all its resources to intra-session network coded flows, as there would be no incentives for the user to request combined packets. In that case, the network users would be unable to benefit from inter-session network coding. On the contrary, by including the delay of the children nodes in the optimization objective, we provide incentives for network nodes to combine packets of different sources in order to serve as many users as possible without major penalty on their own utility. By encouraging nodes' collaboration we reach more socially fair solutions.

Let us denote as ${\bm{r}_i = (r_{ki}^t, r_{ij}^t), \;\forall k\in\mathcal{A}_i,\; \forall j \in \mathcal{D}_i, \;\forall t \in \mathcal{T},}$ the vector of innovative packet flow rates, where $r_{ki}^{t}$ represents the innovative rate of packets of type $t$ received by the node $i$ from its parent $k$, while $r_{ij}^{t}$ is the innovative rate of packets of type $t$ received by the child node $j$ from the node $i$. The distributed delay optimization in the $i$-th node is stated as 
\begin{equation}
	\underset{\bm{r}_{i}}{\operatorname{arg\,min}}\;\overline{\Delta}_i({\bm{r}_{i}}) 
	 \quad \mbox{s.t.} \:\: \bm{r}_i \in\mathcal{R}_{i}^{min}
\label{eq:G}
\end{equation}

The search space $\mathcal{R}_{i}^{min}$ is defined by a set of linear inequality constraints, which determine the set of feasible values of the innovative packet flow rates on the input and output links of the node $i$ 
{
\allowdisplaybreaks
 \begin{flalign}
			&0 \leq \sum\limits_{t\in\mathcal{T}}r_{ki}^t \leq  b_{ki}(1-\pi_{ki}), \quad \forall k\in\mathcal{A}_i \label{eq:X1}\\
			&0 \leq\sum\limits_{t\in\mathcal{T}} r_{ij}^t \leq b_{ij}(1-\pi_{ij}), \quad \forall j\in\mathcal{D}_i \label{eq:X2}\\
%
%
%
%
%
			&\sum\limits_{t^\prime\in\mathcal{T}_{t,s}}r_{ki}^{t^\prime} 
			\leq \sum\limits_{t^\prime \in\mathcal{T}_{t,s}}R_k^{t^\prime},
			\quad \forall t\in\mathcal{T}, \; \; \forall s\in\mathcal{S}_t, \; \; \forall k\in\mathcal{A}_i \label{eq:X7}\\
			&\sum\limits_{t^\prime\in\mathcal{T}_{t,s}}r_{ij}^{t^\prime} 
			\leq \sum\limits_{t^\prime \in\mathcal{T}_{t,s}}\sum\limits_{k\in\mathcal{A}_i}r_{ki}^{t^\prime},
			\quad \forall t\in\mathcal{T}, \; \; \forall s \in\mathcal{S}_t, \; \; \forall j\in\mathcal{D}_i \label{eq:X8}\\
			& \sum\limits_{t^\prime\in\mathcal{T}_{t,s}} \sum\limits_{k\in\mathcal{A}_i}r_{ki}^{t^\prime} \leq U_s,
			\quad \forall  t\in\mathcal{T}, \; \; \forall s\in\mathcal{S}_t \label{eq:X9}\\
			&\sum\limits_{t^\prime\in\mathcal{T}_{t,s}} r_{ij}^{t^\prime} +
			\sum\limits_{t^\prime\in\mathcal{T}_{t,s}} \hat{R}_{j\backslash i}^{t^\prime} \leq U_s ,
			\quad  \forall  t\in\mathcal{T}, \; \;  \forall s\in\mathcal{S}_t,  \; \; \forall j\in\mathcal {D}_i 	\label{eq:X10}	
\end{flalign}
}
The constraints appear in pairs and refer to the input and the output links of the node $i$, respectively. Eqs.~\eqref{eq:X1} and \eqref{eq:X2} are the link capacity constraints, which state that the sum of innovative packet rates for all packet types received on a link cannot exceed the link capacity. Eqs.~\eqref{eq:X7} and \eqref{eq:X8} give upper bounds to the innovative packet flow rates with the available innovative packet rates at parent nodes. Finally, Eqs.~\eqref{eq:X9} and \eqref{eq:X10} limit the innovative packet rate by the available innovative rate provided by the sources, {\em i.e.}, the user cannot receive innovative packets faster than they are injected in the network by the sources. 

The average decoding delay of the node $i$ and its children nodes $\overline{\Delta}_i({\bm{r}_{i}})$ is written as
\begin{equation}
	\overline{\Delta}_i ({\bm{r}_{i}})= \frac{1}{|\mathcal{D}_i|+1}\big(\Delta_i({\bm{r}_{i}}, g_i)+ \sum\limits_{j\in\mathcal{D}_i}\Delta_j({\bm{r}_{i}}, g_j)\big)
\label{eq:I}
\end{equation}
The expected delay $\Delta_i({\bm{r}_{i}}, g_i)$ experienced by the user $i$ for receiving and decoding a block of packets of the requested session $g_i$ depends on the average number of packets that the user $i$ needs to collect for decoding. The latter is a function of the types and the innovative rates of the packets that arrive at the node. 

The optimization problem stated in Eq.~\eqref{eq:G} is complex and in general non-convex. In order to solve it, we make the following simplifying assumptions. We assume that the time is slotted and that at most one packet can be received by the node $i$ in each time slot. We approximate the duration of the time slot by $d_i =\frac{1}{C_i^d}$. Thus, we can estimate the average decoding delay as the product of the average time $d_i$ required to receive one packet and the average number of packets $E[l]$ that the user receives before it is able to decode
\begin{equation}
	\Delta_i({\bm{r}_{i}}, g_i) = d_iE[l]
\label{eq:J}
\end{equation}
The solution of Eq.~\eqref{eq:G} then requires the computation of the average number of packets $E[l]$ that the node and its children nodes need to receive in order to decode their data of interest. Next, we will present an efficient method for computing $E[l]$ that permits to transform the initial problem into a set of convex subproblems and to obtain a solution with low complexity.


\section{Decoding delay analysis with equivalent flows}
\label{sec:equivalentflows}

The estimation of the average decoding delay as described in Eq.~\eqref{eq:J} requires the computation of the expected number of packets $E[l]$ that the node has to receive for decoding one block of packets of the session of interest. The exact computation of $E[l]$ involves considering all the possible events that lead to a decodable set of $l$ packets \cite{BourtsouCentrISNC}. This is clearly non-trivial to compute. In this section, we introduce the notion of equivalent flows in order to approximate the decoding delay with simple functions that can be computed efficiently.

Let us assume that the session $s$ is the session of interest. There are several possibilities to decode the packets of this source data. The session $s$ can be decoded from intra-session network coded packets when $N_{s}$ such innovative packets are available. Otherwise, it can be decoded from a set of intra- and inter-session network coded packets of types $t^{\prime}\in \mathcal{T}_{t}$, for any $t\in \mathcal{T}^{s}$, as long as this set of packets forms a full rank system. In particular, when the decoding is performed from a session combination $t$, one needs $N_{s^{\prime}}$ innovative packets for each component session $s^{\prime}\in \mathcal{S}_{t}$.\footnote {We say that a session $s$ is decoded from the session combination $t\in\mathcal{T}^{s}$, when packets of types $t^{\prime}\in \mathcal{T}_{t}$ are used to decode the data of session $s$.} Note that these $N_{s^{\prime}}$ packets can be of any type $t^{\prime}\in\mathcal{T}_{t,s^{\prime}}$. Note also that any novel inter-session network coded packet of type $t^{\prime}\in\mathcal{T}_{t}$ contains novel information for all the component sessions $s^{\prime}\in\mathcal{S}_{t^{\prime}}$. In other words, any novel inter-session network coded packet of type $t^{\prime}\in\mathcal{T}_{t}$ can increase the rank of any of its component sessions. This property stems from the definition of innovation and is also guaranteed by the constraints \eqref{eq:X7} - \eqref{eq:X10}.

The above observations bring us to the core idea behind the notion of equivalent flows. We can see that, when session $s$ is decoded from the session combination $t$, we can treat every component session $s^{\prime}\in\mathcal{S}_{t}$ of $t$ as a separate session for which we need to collect $N_{s^{\prime}}$ innovative packets of any type $t^{\prime}\in \mathcal{T}_{t,s^{\prime}}$. That means that the flow of packets of type $t^{\prime}$ can be split among its component sessions. The rate at which innovative packets are collected for the component session $s^{\prime}$ is equal to the sum of the contributions of each packet flow $t^{\prime}$ that has $s^{\prime}$ as a component session. The only difference between the decoding of session $s$ from intra-session network coded packets and the decoding of the same session from the session combination $t$ is that, in the latter case, the session $s$ can only be decoded when a sufficient number of innovative packets is available for all  the component sessions $s^{\prime} \in \mathcal{S}_{t}$. We now propose a definition for the \textit{equivalent flows}.
\begin{defn}
Given a session combination $t\in \mathcal{T}$, we define an \textit{equivalent flow} for every component session $s\in\mathcal{S}_{t}$ of $t$ as a virtual flow of packets with innovative rate equal to the sum of the contributions of innovative rates from every flow of type $t^{\prime}\in \mathcal{T}_{t,s}$.
\label{def:D1}
\end{defn}

We will henceforth refer to the rate of an equivalent flow as \textit{equivalent rate}. Note that Definition \ref{def:D1} is general and applies also to types $t$ that correspond to intra-session network coded packets. In this case, the equivalent flow coincides with the actual innovative flow of intra-session network coded packets. When $t$ is a combination of two or more sessions, the innovative rate of the equivalent flow for every component session $s\in\mathcal{S}_{t}$ is higher or equal to the actual innovative rate of the flow of intra-session network coded packets of this same component session. This increment in the equivalent rate comes from the contribution of the inter-session network coded packet flows that have the session $s$ as a  component session.

Mathematically, the equivalent innovative rate $v_{i,s}^{t}$ for the component session $s$ in the session combination $t$  received at the node $i$ can be represented as
\begin{equation}
  	v_{i,s}^{t} = \sum_{t^{\prime}\in\mathcal{T}_{t,s}}\gamma_{i,s}^{t,t^{\prime}}\sum_{k\in\mathcal{A}_{i}}r_{ki}^{t^{\prime}}, \quad \forall s\in\mathcal{S}_{t}
\label{eq:F0}
\end{equation}
where $ \gamma_{i,s}^{t,t^{\prime}} \in [0,1]$ and $\sum_{s^{\prime}\in\mathcal{S}_{t^{\prime}}}\gamma_{i,s^{\prime}}^{t,t^{\prime}} \leq 1$, $\forall t^{\prime} \in \mathcal{T}_{t}$. The coefficient $\gamma_{i,s}^{t,t^{\prime}}$ indicates the contribution of the innovative flow of type $t^{\prime}$ to the rate $v_{i,s}^{t}$  at which innovative packets are collected for the component session $s$ when session combination $t$ is considered for decoding.

Equipped with the definition of equivalent flows, we can now calculate the coefficients $\gamma_{i,s}^{t,t^{\prime}}$ and approximate the decoding delay at node $i$. Let us denote as 
\begin{equation}
	p_i^t = \frac{\sum_{k\in\mathcal{A}_i}r_{ki}^t}{C_i^d}
\label{eq:F6}
\end{equation}
the probability of receiving an innovative packet of type $t$ at node $i$, where $r_{ki}^t$ is the innovative flow rate of packets of type $t$ that the node $i$ receives from its parent $k$. In a similar way, for a given session combination $t\in\mathcal{T}$ and for every component session $s\in\mathcal{S}_{t}$, we define the probability
\begin{equation}
	q_{i,{s}}^t = \frac{v_{i,s}^{t}}{C_i^d} = \frac{\sum_{t^{\prime}\in\mathcal{T}_{t,s}}\gamma_{i,s}^{t,t^{\prime}}\sum_{k\in\mathcal{A}_{i}}r_{ki}^{t^{\prime}}}	{C_i^d} = \sum_{t^{\prime}\in\mathcal{T}_{t,s}}\gamma_{i,s}^{t,t^{\prime}}p_i^{t^{\prime}}
\label{eq:F7}
\end{equation}
which represents the probability of receiving an innovative packet for the component session $s$ at node $i$ assuming that decoding is performed from the session combination $t$. The probability $P_{i,{s}}^t(l) $ to receive the $N_{s}$-th innovative packet for the component session $s$ of the session combination $t$ upon receiving exactly $l$ packets at node $i$ is given by the negative binomial distribution
\begin{equation}
	P_{i,{s}}^t(l) = \binom{l-1}{N_{s} - 1}(q_{i,s}^t)^{N_{s}}(1-q_{i,s}^t)^{l-N_{s}}
\label{eq:F1}
\end{equation}
Thus, the average time needed for receiving $N_{s}$ innovative packets for the component session $s$ at node $i$ is
\begin{equation}
	\Delta_{i,s}^{t} = d_{i}E[l] = d_{i} \sum_{l= N_{s}}^\infty l P_{i,{s}}^t(l) = d_{i}\frac{N_{s}}{q_{i,s}^t}
\label{eq:F2}
\end{equation}
where $E[l]$ stands for the average number of packets that the user has to receive in order to collect $N_{s}$ innovative packets for the component session $s$. It is given by the mean of the negative binomial distribution in Eq.~\eqref{eq:F1} and, in our case, it is simply the ratio of the size of the block of source packets $N_{s}$ and the probability ${q_{i,s}^t}$ of receiving an innovative packet. Note that, in Eqs.~\eqref{eq:F1} and \eqref{eq:F2} we have assumed that the innovative rate is independent of the number of packets stored in the node's buffer. In practice, as the number of innovative packets in the node's buffer increases, the probability of receiving a non-innovative packet also increases. However, in networks with high path diversity and for large Galois field sizes the probability of generating two identical or linearly dependent packets is negligible \cite{RandomizedNC03}. This permits to make the assumption that the innovative rate does not depend on the number of packets stored in the node's buffer. 

In order to determine the values of the $\gamma_{i,s}^{t,t^{\prime}}$ coefficients for every session combination $t$ we need to look at the problem from the point of view of the decoder. The decoding of the session $s$ from a session combination $t$ is feasible as soon as $N_{s^{\prime}}$ innovative packets of any type $t^{\prime}\in\mathcal{T}_{t,s^{\prime}}$ are available at the decoder for every component session $s^{\prime} \in \mathcal{S}_{t}$. This implies that the inter-session network coded flows are split among their component sessions in such a way that the delays for collecting the necessary number of innovative packets for every component session are as balanced as possible. That means that the equivalent rates, as seen by the decoder, are such that the maximum of the delays among the component sessions is minimized.

We can now formulate the $ \min\max$ optimization problem that permits to determine the coefficients $\gamma_{i,s}^{t,t^{\prime}}$  and subsequently the equivalent rates. The objective is to calculate the coefficients ${\bm \gamma}_{i}^{t} =\{\gamma_{i,s}^{t,t^{\prime}}\}$ that minimize the maximum average delay $\Delta_{i,s}^{t}$ among the component sessions $s\in\mathcal{S}_{t}$.  Formally, this optimization problem is written as
\begin{equation}
	\begin{split}
		&\min _{{\bm \gamma}_{i}^{t}} \max _{s\in\mathcal{S}_t}\Delta_{i,s}^{t}({\bm \gamma}_{i}^{t}) = \min _{{\bm \gamma}_{i}^{t}} \max _{s\in\mathcal{S}_t}d_{i}\frac{N_{s}}{q_{i,s}^t({\bm \gamma}_{i}^{t})} \\
	& \mbox{s.t.} \sum_{s^\prime \in \mathcal{S}_{t^\prime}}{\gamma_{i,{s^\prime}}^{t,t^\prime}} \leq 1, \gamma_{i,{s^\prime}}^{t,t^\prime} \in [0,1] , \; \forall t^\prime \in \mathcal{T}_{t}
	\end{split}
	 \label{eq:F3}
\end{equation}

Once we have computed the equivalent rates, we can estimate the average decoding delay $\Delta_{i}^{t}$ experienced by the user $i$ for decoding a block of packets of session $s$ from the session combination $t$. Assuming that $\hat{\bm{\gamma}}_{i}^{t}$ is the optimal solution of the optimization problem in Eq.~\eqref{eq:F3}, the decoding delay $\Delta_{i}^{t}$ is simply the maximum of delays $\Delta_{i,s}^{t}$ over all the component sessions $s\in\mathcal{S}_{t}$. Indeed, in order to decode session $s$ the user needs to wait until the necessary number of packets is available for every component session. Thus, we have
\begin{equation}
	\Delta_{i}^{t} = \max_{s\in\mathcal{S}_t}\Delta_{i,s}^{t}(\hat{\bm{\gamma}}_{i}^{t})
\label{eq:F4}
\end{equation}
Note that the vector of coefficients $\hat{\bm{\gamma}}_{i}^{t}$ is different for different session combinations $t$.

To complete our analysis, we need to determine the average decoding delay observed at node $i$ for decoding its session of interest $s$. Eq.~\eqref{eq:F4} gives the average decoding delay under the assumption that the user decodes from the specific session combination $t$ that has $s$ as a component session. In general, there may be multiple session combinations $t^{\prime}$ such that $s\cap\mathcal{S}_{t^{\prime}} \neq\emptyset$ yielding thus several possibilities for decoding. However, to simplify our analysis, we will assume that, for a given set of innovative packet flow rates on the user's input links, the user decodes the data of interest from the session combination that corresponds to the minimum average decoding delay $\Delta_{i}^{t}$
\begin{equation}
	\Delta_{i}({\bm{r}_{i}}, s) = \min_{t \in \mathcal{T}^{s} } \Delta_{i}^{t} =  \min_{t \in \mathcal{T}^{s} }  \max_{s^{\prime}\in\mathcal{S}_t}\Delta_{i,s^{\prime}}^{t}(\hat{\bm{\gamma}}_{i}^{t})
\label{eq:F5}
\end{equation}

To summarize, in order to compute the approximate decoding delay for decoding packets of session $s$ we have the following steps:
 \begin{enumerate}
 \item we compute the equivalent flows by solving the $\min \max$ problem in Eq.~\eqref{eq:F3} for every session combination $t\in \mathcal{T}^{s}$,
 \item using the equivalent flows computed in step 1, we calculate the approximate decoding delay for every session combination $t$ from Eqs.~\eqref{eq:F2} and \eqref{eq:F4},
 \item finally, we approximate the delay with the minimum among the delays computed in step 2 (Eq.~\eqref{eq:F5}).
 \end{enumerate}

Finally, it should be noted that we have considered the worst case scenario where all the component sessions involved in a session combination have to be decoded along with the requested session. This is due to the random encoding strategy deployed in our scheme. Other encoding strategies could be devised to avoid decoding all the sessions \cite{MarkInterfAlign}. However, these strategies require expensive control and diminish the advantages of randomized network coding. The design of such encoding strategies is not trivial and is out of the scope of this paper. 

We now illustrate the computation of  equivalent flows and the estimation of the decoding delay with a numerical example.

\begin{example}
\label{ex:E1_ch2}

We assume that three sources, namely $s_{1}, s_{2}$ and $s_{3}$, are transmitted into the network and that the user $i$ requests the session $s_1$. The block sizes for the three sessions are $N_{s_1} = N_{s_2}=N_{s_3} = 10$ packets. We choose two sets of probabilities of receiving an innovative packet at the node $i$ for all the possible packet types $t\in\mathcal{T}$. These two sets of probabilities, shown in Table \ref{tab:actualvalues}, represent two different rate allocations at the node $i$ and correspond to two different instances of the decoding problem, namely Problem A and Problem B. Given these probabilities, we want to estimate the decoding delay at the node $i$ for decoding one block of packets from its source of interest. The session $s_{1}$ can be decoded from any of the session combinations $t = s_{1}, t= s_{1}s_{2}, t = s_{1}s_{3}$ or $t = s_{1}s_{2}s_{3}$.

\begin{table*}[t]
	\begin{center}
	\caption{Probabilities $p_{i}^{t}$ of receiving an innovative packet of type $t$ at the node $i$ for all the possible packet types $t \in \mathcal{T}$ in the Example \ref{ex:E1_ch2}.\label{tab:actualvalues}}
		\begin{tabular}{| c || c| c | c | c | c | c | c |}
		\hline 
		& $ p_i^{s_1}$ & $ p_i^{s_2}$  & $ p_i^{s_3}$& $ p_i^{s_1s_2}$& $p_i^{s_1s_3}$& $ p_i^{s_2s_3}$& $ p_i^{s_1s_2s_3}$\\
		\hline 
		\textbf {Problem A} & 0.1824 & 0.2022 & 0.2035 & 0.0385 & 0.1439 & 0.0323 & 0.0707\\
		\hline
		\textbf{Problem B} & 0.0556 & 0.0278 & 0.2778 & 0.1111 & 0.0833 & 0.3889 & 0.0111\\
		\hline
		\end{tabular}
	\end{center}
\end{table*}

\begin{table}[t]
	\begin{center}
	\caption{Probabilities $q_{i,s}^{t}$ associated with the equivalent rates for all possible session combinations $t \in \mathcal{T}^{{s_{1}}}$ in the Example \ref{ex:E1_ch2}.\label{tab:equivalentvalues}}
		\begin{tabular}{ c  | c |  c | c | c | c | c | c | c |}
		\cline{1-9}
		\multicolumn{1}{|c||}{}  &  $t = s_{1}$  &  \multicolumn{2}{ c |}{$t = s_{1}s_{2}$}  &  \multicolumn{2}{ c |}{$t = s_{1}s_{3}$}  &  \multicolumn{3}{c |}{$t = s_{1}s_{2}s_{3}$} \\
		\cline{2-9}
		\multicolumn{1}{|c||}{}  &  $q_{i,{s_1}}^{s_{1}}$  &  $q_{i,{s_1}}^{s_{1}s_{2}}$  &  $q_{i,{s_2}}^{s_{1}s_{2}}$  &  $q_{i,{s_1}}^{s_{1}s_{3}}$  &  $q_{i,{s_3}}^{s_{1}s_{3}}$  &  $q_{i,{s_1}}^{s_{1}s_{2}s_{3}}$  &  $ q_{i,{s_2}}^{s_{1}s_{2}s_{3}}$  &  $ q_{i,{s_3}}^{s_{1}s_{2}s_{3}}$\\
		\hline
		 \multicolumn{1}{|c||}{\textbf {Problem A} }  & 0.1824  &  0.2116  &  0.2116  &  0.2649  &  0.2649  &  0.2912  &  0.2912  &  0.2912 \\
		\hline
		 \multicolumn{1}{|c||}{\textbf {Problem B} }  & 0.0556  &  0.0972  &  0.0973  &  0.1389  &  0.2778  &  0.2611  &  0.3473  & 0.3473 \\
		\hline
		\end{tabular}
	\end{center}
\end{table}

\begin{table}[t!]
	\begin{center}
	\caption{The average number of packets $E_{i}^{t}$ required at node $i$ in order to decode the session of interest from the session combination $t$ in the Example \ref{ex:E1_ch2}.\label{tab:avgpackets}}
		\begin{tabular}{ | c || c | c | c | c | c | }
		\hline
		& $E_{i}$  &  $E_{i}^{s_{1}}$  &  $E_{i}^{s_{1}s_{2}}$  &  $E_{i}^{s_{1}s_{3}}$  & $E_{i}^{s_{1}s_{2}s_{3}}$\\
		\hline
		 {\textbf {Problem A} }& 33.8 & 54.8  & 47.3  & 37.6  & \textbf{34.3}  \\
		\hline
		{\textbf {Problem B} } & 39.7  & 179.9 & 102.9  & 72.0 & \textbf{38.3} \\
		\hline
		\end{tabular}
	\end{center}
\end{table}

Table \ref{tab:equivalentvalues} illustrates the results obtained by following the three steps summarized above. In particular, in Table \ref{tab:equivalentvalues} we present the probabilities $q_{i,s}^{t}$ that correspond to the equivalent flows for all possible session combinations that have session $s_{1}$ as a component session. In Table \ref{tab:avgpackets} we also present the average number of packets $E_{i}^{t}$ that have to be received by the user $i$ in order to decode its session of interest from the session combination $t$. Finally, for comparison, we compute the average number of packets $E_{i}$ required for decoding using the method provided in \cite{BourtsouCentrISNC}.

\begin{figure*}[t]
	\begin{center}
		~\includegraphics[width=0.89\textwidth]{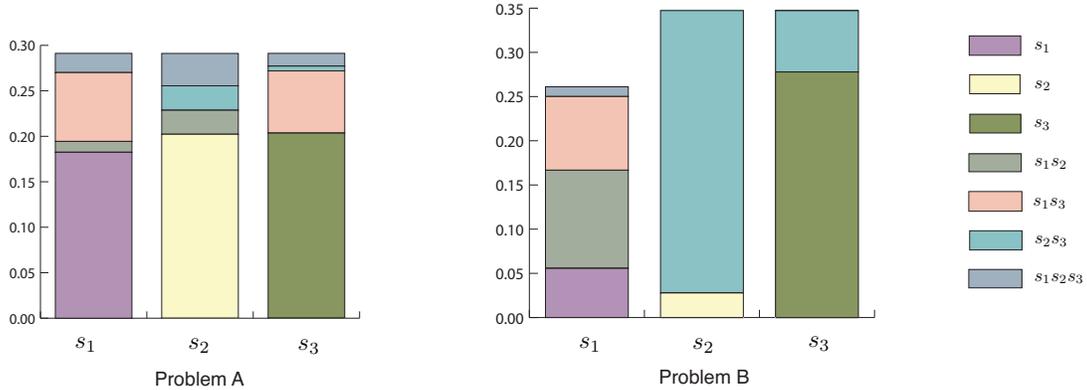}
	\end{center}
	\caption{Illustration of the contribution of each packet flow to the equivalent rates of the component sessions $s_{1}$, $s_{2}$ and $s_{3}$, when session combination $t = s_{1}s_{2}s_{3}$ is considered for decoding. \label{fig:equivalentflows}}
\end{figure*}

From the results presented in Table \ref{tab:equivalentvalues}, we can see that, when the session $s_{1}$ is decoded from intra-session network coded packets, the equivalent rate is equal to the actual innovative rate of intra-session network coded packets of type $t = s_{1}$, {\em i.e.}, $q_{i,s_{1}}^{s_{1}}= p_{i}^{s_{1}}$. On the other hand, when an inter-session combination is considered for decoding, the equivalent rates of the component sessions are higher than the innovative rates of the intra-session network coded flows of the component sessions. For example, when the session combination $t=s_{1}s_{3}$ is considered for decoding, we have $q_{i,s_{1}}^{s_{1}s_{3}} > p_{i}^{s_{1}}$ and $q_{i,s_{3}}^{s_{1}s_{3}} > p_{i}^{s_{3}}$. The increment in the rate comes from the splitting of the combined flow of type $t = s_{1}s_{3}$ among its component sessions $s_{1}$ and $s_{3}$. Further, according to the results presented in Table \ref{tab:avgpackets}, we can observe that in both Problems A and B the minimum number of packets required for decoding corresponds to the session combination $t = s_{1}s_{2}s_{3}$. We can also see that this number, calculated using the approach of equivalent flows, is very close to the actual average number of packets computed with the method provided in \cite{BourtsouCentrISNC}. Another observation that we can make is that the performance in terms of decoding delay, for a given session combination, is driven by the component session that requires the most time to collect all the necessary innovative packets. Let us consider again the session combination $t = s_{1}s_{3}$. The equivalent rates for the component session $s_{3}$ are almost the same in both Problems A and B, however, the equivalent rate for the component session $s_{1}$ in Problem B is approximately half of the corresponding rate in Problem A and also half of the equivalent rate for the component session $s_{3}$. Thus, in Problem B the user needs to collect approximately two times more packets than in Problem A, in order to decode from the session combination $t = s_{1}s_{3}$. 

Finally, Fig.~\ref{fig:equivalentflows} illustrates the contributions of each packet flow to the equivalent rates of the component sessions $s_{1}$, $s_{2}$ and $s_{3}$, when the session combination $t = s_{1}s_{2}s_{3}$ is considered for decoding. Every color corresponds to a specific packet flow type. The bars represent the equivalent rates. The height of each bar is proportional to the magnitude of the corresponding equivalent rate. The height of a sub-bar of a certain color is proportional to the contribution of the flow denoted with the same color. We can see that, in Problem A, the packet flows are split among their component sessions in such a way that the equivalent rates for all component sessions are equal. On the contrary, in Problem B, we see that all the flows that have session $s_{1}$ as a component session contribute only to the equivalent rate that corresponds to $s_{1}$. Furthermore, this equivalent rate is lower than the ones that correspond to the component sessions $s_{2}$ and $s_{3}$.    
\end{example}


\section{Distributed rate allocation}
\label{sec:ratealloc}

In this section we present the distributed rate allocation algorithm that solves the problem stated in Eq.~\eqref{eq:G} with the help of the equivalent flow representation. According to our proposed solution, every node solves a rate allocation optimization problem in two steps. In every optimization round, the node first finds the optimal rate allocation that improves the average decoding delay for itself and its direct children. In order to find the optimal rate allocation, the initial problem is decomposed into several convex subproblems based on the equivalent flows representation described in Section \ref{sec:equivalentflows}. Second, the node maximizes the total throughput in terms of innovative packet rate while preserving the optimal rates obtained from the delay minimization step. This second step compensates for the partially myopic behavior of the network nodes and boosts the performance of the data delivery system as each user can transmit packets that are potentially useful for other users different than its direct children.


\subsection{Decoding delay minimization}
\label{sec:delaymin}

The first step of our algorithm consists in finding the rates that minimize the decoding delay of a node and its direct children. In order to determine these rates, the network node first obtains all the necessary information from its neighborhood following the communication protocol described in Section \ref{sec:protocol}. It then solves the rate allocation problem independently of the other network nodes and without any centralized control. 

The decoding delay minimization problem is stated in Eq.~\eqref{eq:G}. Recall that we have made a simplifying assumption that, for a given rate allocation, the network user $i$ and its direct children $j$, $(j \in \mathcal{D}_{j})$, decode the requested data from the session combinations that correspond to the minimum decoding delay (see Eq.~\eqref{eq:F5}). Hence, the original problem can be decomposed into a set of convex subproblems. Every subproblem corresponds to finding the optimal rate allocation vector  $\bm{r}_i = (r_{ki}^t, r_{ij}^t), \forall k\in\mathcal{A}_i, \forall j \in \mathcal{D}_i, \forall t \in \mathcal{T},$ that yields the minimum average decoding delay $\overline{\Delta}_i({\bm{r}_{i}})$ for a specific tuple of session combinations $(t_i, \{t_j,j\in\mathcal{D}_i\})\in \mathcal{T}^{g_{i}}\times\underset{ j \in \mathcal{D}_i}{\prod}\mathcal{T}^{g_{j}}$. Combining Eqs.~\eqref{eq:G}, \eqref{eq:I}, \eqref{eq:F4} and \eqref{eq:F2}, the subproblem of finding the optimal rate allocation for a given tuple  $(t_i, \{t_j,j\in\mathcal{D}_i\})\in \mathcal{T}^{g_{i}}\times\underset{ j \in \mathcal{D}_i}{\prod}\mathcal{T}^{g_{j}}$, can be written as 
\begin{equation}
\allowdisplaybreaks
\begin{split}
	\underset{\bm{r}_{i}, {\bm \gamma}_{i}^{t_{i}},{\bm \gamma}_{j}^{t_{j}}}{\operatorname{arg\,min}} &\; \frac{1}{|\mathcal{D}_i|+1}(d_i \max_{s \in \mathcal{S}_{t_i}} \frac{N_{s}}{q_{i,s}^{t_i}({\bm \gamma}_i^{t_{i}})}
	+ \sum\limits_{j\in\mathcal{D}_i}d_j \max_{s \in \mathcal{S}_{t_j}} \frac{N_{s}}{q_{j,s}^{t_j}({\bm \gamma}_j^{t_{j}})})  \\
	 \mbox{s.t.} &\quad \bm{r}_i \in  \mathcal{R}_{i}^{min} \\
	& \sum_{s^\prime \in \mathcal{S}_{t^{\prime}}}{\gamma_{n,{s^\prime}}^{t_{n},t^{\prime}}}  \leq 1, \; \forall t^{\prime} \in \mathcal{T}_{t_{n}},\; \forall n\in \{i \cup \mathcal{D}_i \}
\end{split}
\label{eq:H1}
\end{equation}
From Eq.~\eqref{eq:F7}, $q_{i,{s}}^t =  \sum_{t^{\prime}\in\mathcal{T}_{t,s}}\gamma_{i,s}^{t,t^{\prime}}p_i^{t^{\prime}}$. We can replace the product $\gamma_{i,s}^{t,t^{\prime}}p_i^{t^{\prime}}$ with the variable $x_{i,s}^{t,t^{\prime}}\in [0,1]$ and write $q_{i,{s}}^t =  \sum_{t^{\prime}\in\mathcal{T}_{t,s}}x_{i,s}^{t,t^{\prime}}$. The minimization problem in Eq.~\eqref{eq:H1} becomes
\begin{equation}
{
\allowdisplaybreaks
\begin{split}
	\underset{\bm{r}_{i}, {\bm x}_{i}^{t_{i}},{\bm x}_{j}^{t_{j}}}{\operatorname{arg\,min}} &\; \frac{1}{|\mathcal{D}_i|+1}(d_i \max_{s \in \mathcal{S}_{t_i}} \frac{N_{s}}{q_{i,s}^{t_i}({\bm x}_i^{t_{i}})}
	+ \sum\limits_{j\in\mathcal{D}_i}d_j \max_{s \in \mathcal{S}_{t_j}} \frac{N_{s}}{q_{j,s}^{t_j}({\bm x}_j^{t_{j}})})  \\
	 \mbox{s.t.} &\quad \bm{r}_i \in \mathcal{R}_{i}^{min} \\
	& \sum_{s^\prime \in \mathcal{S}_{t^{\prime}}}{x_{n,{s^\prime}}^{t_{n},t^{\prime}}}  \leq p_{i}^{t^{\prime}}, \; \forall t^{\prime} \in \mathcal{T}_{t_{n}},\; \forall n\in \{i \cup \mathcal{D}_i \}
\end{split}
}
\label{eq:H2}
\end{equation}
where $p_i^t = \frac{\sum_{k\in\mathcal{A}_i}r_{ki}^t}{C_i^d}$ and $p_j^t = \frac{\hat{R}_{j\backslash i}^t +r_{ij}^t}{C_j^d},j\in\mathcal{D}_i$. The problem stated in Eq.~\eqref{eq:H2} is convex. This can be shown using the following arguments. The function $\frac{N_{s}}{q_{i,s}^{t_i}(\bm{x}_i^{t_{i}})}$ is convex since it is a composition of the convex function $\frac{1}{x}$ with the affine expression $q_{i,s}^{t_i}(\bm{x}_i^{t_{i}})$. The pointwise maximum is also a convex function. Thus, the objective function in Eq.~\eqref{eq:H1} is convex since it is a nonnegative weighted sum of convex functions \cite{convopt}. It can be solved using the CVX Matlab-based package \cite{cvx} for example.

The solution to the initial rate allocation problem stated in Eq.~\eqref{eq:G} can be obtained by solving the subproblems of Eq.~\eqref{eq:H2} for all the feasible tuples $(t_i, \{t_j,j\in\mathcal{D}_i\})\in \mathcal{T}^{g_i}\times\underset{ j \in \mathcal{D}_i}{\prod}\mathcal{T}^{g_j}$. The results of these subproblems are then combined and the solution (rate allocation vector) that yields the minimum delay is chosen. This solution also constitutes the solution to the original problem in Eq.~\eqref{eq:G}. The number of convex subproblems to be solved depends on the cardinality of the set $ \mathcal{T}^{g_i}\times\underset{ j \in \mathcal{D}_i}{\prod}\mathcal{T}^{g_j}$, that grows exponentially with the number of sources available in the network and the number of the node's direct children. In practice, however, the number of sources is typically small, and the network users have a limited upload bandwidth, which allows only a few children nodes to be connected simultaneously to the same node. Therefore, the number of convex subproblems to be solved by each node is typically small.


\subsection{Maximization of the total innovative input rate}
\label{sec:ratemax}

The solution of the minimization problem in Eq.~\eqref{eq:G} guarantees the transmission of data sources that are requested by the node $i$ and its children nodes at optimal rates, as long as these data sources are available at the parents of the examined node. However, a node is not aware of the data requested by users that are two or more hops away. This means that, in certain cases, some of the sessions that are available in the network and may be potentially useful for other users beyond the node's $i$ neighborhood, are never requested by the node $i$. Thus, these sessions can never be forwarded when requested by other network users, which eventually penalizes the performance of the downstream nodes. This drawback of the distributed scheme is a result of having a limited network horizon with only local information in solving the rate allocation. In order to reduce the effect of this shortcoming, we propose to solve a simple throughput maximization problem. This maximization problem is solved in every optimization round immediately after the optimal rates have been determined as presented in Section \ref{sec:delaymin}. Specifically, we aim at maximizing the total innovative packet flow rate for all the packet types such that the flow values are larger or equal to the optimal flow rates computed from Eq.~\eqref{eq:G}. Practically, this means that, whenever there exists some unused bandwidth, it is allocated to packet flows that are not explicitly requested by a node or its children nodes, but that can be potentially useful for other nodes. The maximization problem can be formally written as
\begin{equation}
\begin{split}
			\underset{\bm{r}_i}
			{\operatorname{arg\,max}}\;\sum\limits_{ k\in\mathcal{A}_i}\sum\limits_{t\in\mathcal{T}}r_{ki}^t 
			  \quad \mbox{s.t.} \:\: \bm{r}_i \in\mathcal{R}_{i}^{max} \;\; \mbox{and} \;\; \bm{r}_i \geq \bm{r}^{min}_{i}
\end{split}
\label{eq:L}
\end{equation}
where $\bm{r}^{min}_{i} = \underset{\bm{r}_{i}}{\operatorname{arg\,min}}\;\overline{\Delta}_i(\bm{r}_{i})$ and the inequality sign between two vectors denotes the inequality relationship between vector elements at the same positions. The search space $\mathcal{R}_{i}^{max}$ is defined by linear inequality constraints given in Eqs.~\eqref{eq:X1}, \eqref{eq:X7} and \eqref{eq:X9}. The optimization problem stated in Eq.~\eqref{eq:L} is a linear program and can be solved using any of the standard optimization algorithms \cite{convopt}.

Note that, since $r^t_{ki}$ is the rate at which innovative packets of type $t$ arrive at node $i$ from its parent node $k$, the actual rate $f_{ki}^t$  that the node $i$ has to request from its parent should be augmented by the average packet loss rate that is observed on the link 
\begin{equation}
	f_{ki}^t = \frac{r_{ki}^t}{1-\pi _{ki}}
	\label{eq:M1}
\end{equation} 

\floatname{algorithm}{Algorithm}
\begin{algorithm}[t]
	\caption{Distributed Rate Allocation Algorithm.} 
	\label{algo:optimization}
	\begin{algorithmic}[1] 

	\STATE \textbf{Initialization}

	\STATE Set the current optimization round $l_i=0$ ($l_i = \infty$ for source nodes).
	
	\STATE Define the maximum number of optimization rounds $l_{max}$, the minimum number of optimization rounds $l_{min}$ and the number of optimization rounds $l_{s}$.

	\WHILE{$l_i < l_{max}$}
		
		\STATE  Request the values $l_k$, $ \forall k\in\mathcal{A}_i$, from the parent nodes.
		
		\IF {$l_k > l_i$  $\forall k\in\mathcal{A}_i$}
			
			\STATE Request the values of $\bm{R}_k$, $\forall k\in\mathcal{A}_i$, from the parent nodes
		 			and the values $\hat{\bm{R}}_{j\backslash i}$, $g_j$ and $C_j^d$, $\forall j\in\mathcal{D}_i$, 
					from the children nodes.
			
			\STATE Solve the delay minimization subproblem (Eq.~\eqref{eq:H2}) $\forall (t_i, t_{j,{j\in\mathcal{D}_i}}) 
					\in \mathcal{T}^{g_i}\times\underset{ j \in \mathcal{D}_i}{\prod}\mathcal{T}^{g_j}$. Combine the results and 
					determine the optimal rate allocation vector ${\bm{r}_{i}} = (r_{ki}, r_{ij}),\forall k\in\mathcal{A}_i, \; \forall j\in\mathcal{D}_i$.
			
			\STATE Solve the throughput maximization problem (Eq.~\eqref{eq:L}) and update the rates $r_{ki}^t , \forall k\in\mathcal{A}_i$
			
			\STATE Compute the actual rates $f_{ki}^t, \forall k \in \mathcal{A}_i$ to be requested from the parent nodes, as $f_{ki}^{t} = \frac{r_{ki}^{t}}{1-\pi_{ki}}$
			
			\IF   {$l_k == \infty \; \forall k\in\mathcal{A}_i$ and $l_i > l_{min}$ and $\Delta_i({\bm r}, g_i)$ has not changed for $l_s$ rounds}
				
				\STATE Set $l_i= \infty$
				
			\ELSE 
				
				\STATE Set $l_i = l_i + 1$.
				
			\ENDIF
		
		\ELSE
		
			\STATE Go to step 4.
		
		\ENDIF
	
	\ENDWHILE
	
\end{algorithmic}
\end{algorithm}

The communication protocol and the distributed rate allocation algorithm are summarized in Algorithm \ref{algo:optimization}. The algorithm runs periodically in every network node. This allows to adapt the rate allocation to possible changes that may occur in the network. In practice, a node optimizes its input rates only when all its parent nodes have also performed the optimization. The optimization stops when the utility of the user does not change for a certain number of optimization rounds and all its parents have stopped optimizing or if  a maximum number of optimization rounds has been reached.


\section{Delay performance evaluation}
\label{sec:evaluation}

In this section, we evaluate the performance of the proposed scheme in terms of the average decoding delay. The decoding delay is measured as the time needed for a network node to collect and decode one block of packets from the source of interest. First, we provide an in-depth study of the behaviour of our rate allocation scheme in a small size toy network. We then present the result of applying the proposed method to larger topologies. We compare the performance of our scheme, henceforth denoted as ``InterNC'' (Inter-session Network Coding) to a baseline intra-session network coding rate allocation scheme ``IntraNC'' (Intra-session Network Coding). The latter is a modification of the proposed method except for the fact that the coding across different sessions in the network nodes is not allowed. For the sake of completeness, we also provide a comparison of the decoding delay and the optimal rate allocation with a centralized algorithm that solves the rate allocation problem for the whole network. Note that the centralized scheme assumes full knowledge of the network statistics, and has a complexity that grows exponentially with the number of nodes in the network, so that it does not represent a viable solution in large networks. 


\subsection{Toy network}
\label{sec:toynetwork}

\begin{figure}[t!]
	\begin{center}
		\subfloat[]{\label{fig:toynetwork}\includegraphics[width=0.45\textwidth]{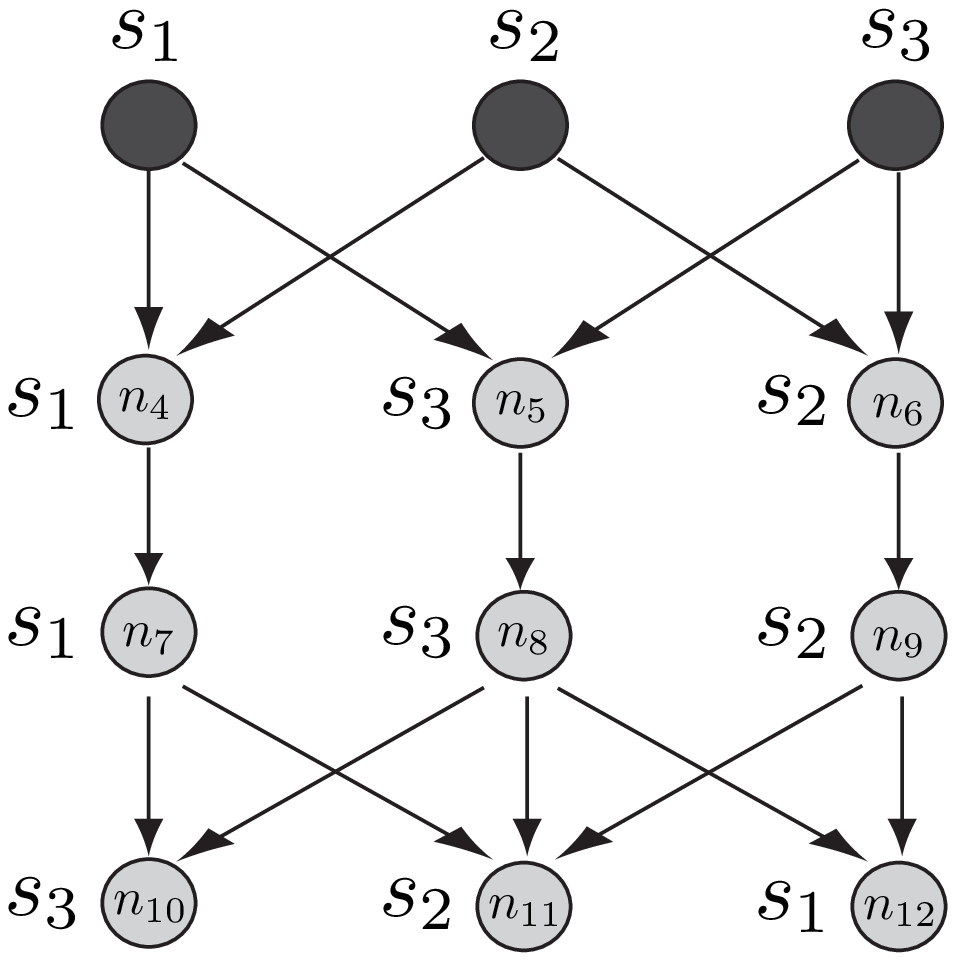}} 
		\subfloat[]{\label{fig:delay_toynetwork}\includegraphics[width=0.53\textwidth]{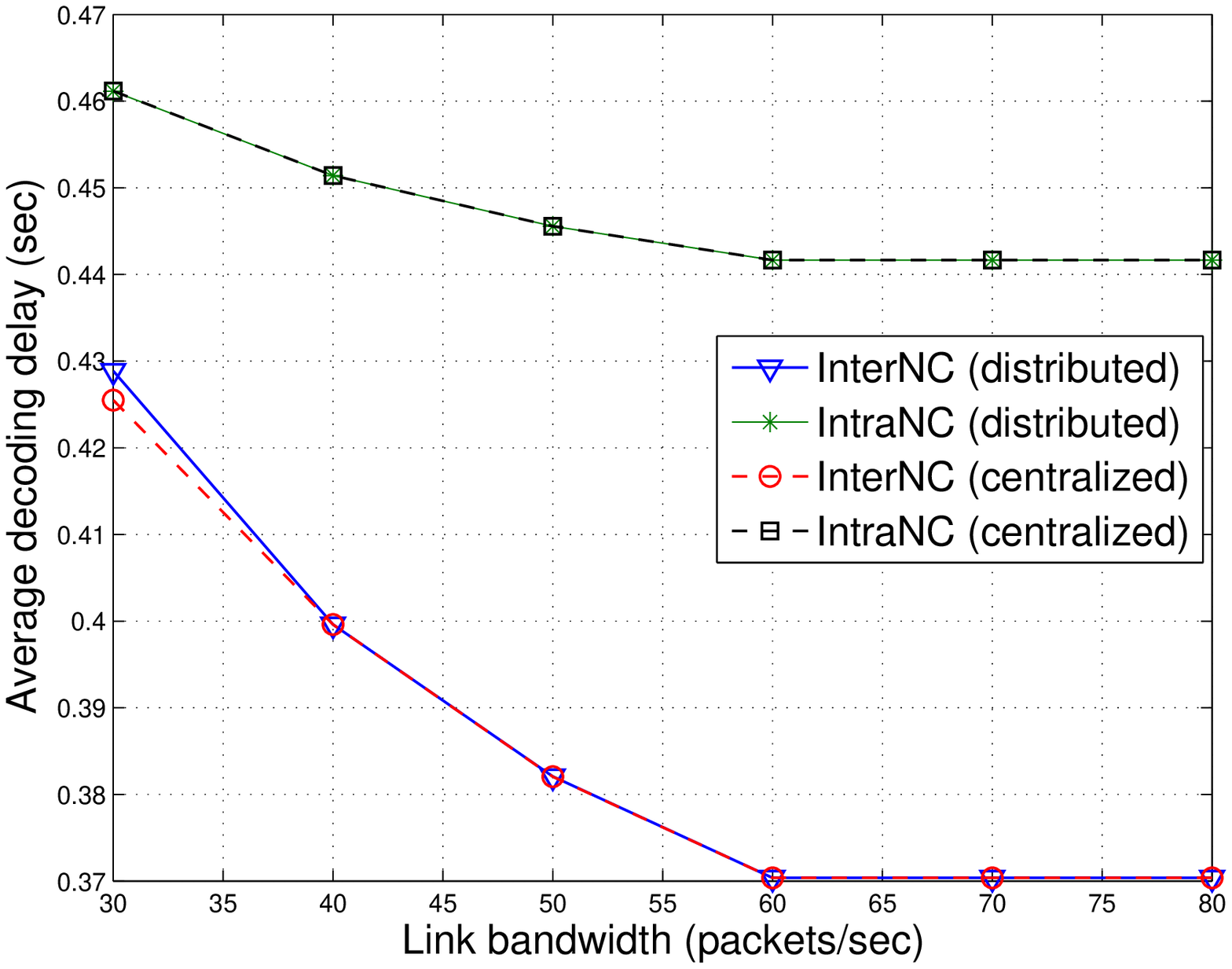}}
	\end{center}
	\caption{(a) Toy P2P network topology where 3 data sources are concurrently transmitted to the network users. The source label next to each node indicates the source data that this node wants to receive. (b) Average decoding delay for the toy network topology depicted in Fig.~\ref{fig:topology1}(a) versus the bandwidth of the links connecting nodes $n_4,n_7$ and $n_6,n_9$. \label{fig:topology1} }	
\end{figure}

We first evaluate the performance of the proposed distributed inter-session rate allocation algorithm for the network depicted in Fig.~\ref{fig:topology1}(a). The network consists of 3 sources and 9 users, which subscribe to different sources. The packet loss rate is set to 5\% on all links. The bandwidth of the links that originate from the sources, as well as of the link connecting nodes $n_5$ and $n_8$ is set to 30 packets/sec. The bandwidth of the links that originate from nodes $n_7, n_8$ and $n_9$ is set to 60 packets/sec. The block size for all 3 sources is 10 packets.

Fig.~\ref{fig:topology1}(b) presents  the evolution of the average delay of the network clients with respect to the bandwidth of the links connecting nodes $n_4,n_7$ and $n_6,n_9$ for all the schemes under comparison. We can observe that, even for low link rates, the proposed distributed InterNC rate allocation scheme performs better than the distributed IntraNC scheme. The gains come from the fact that the nodes can combine packets from different sessions on bottleneck links, whereas in intra-session network coding the performance is limited by the presence of low rate links that cannot serve all the clients at the same time. As the link rates increase, higher gains in terms of delay can be noticed for our proposed InterNC scheme, as more packets are combined across different sessions. On the contrary, the IntraNC schemes fail to deal efficiently with the bottleneck created on the link between the nodes $n_5$ and $n_8$ and the slight improvement of the average decoding delay comes only from the increase of the rate at which packets are supplied to node $n_{11}$. 

Finally, we can notice that the distributed rate allocation schemes, both the proposed InterNC scheme and the baseline IntraNC network coding scheme, manage to reach the performance of their centralized counterpart. This essentially means that for this specific network topology the limited knowledge of the local network statistics that is available to the distributed rate allocation algorithms is sufficient to achieve the global optimal rate allocation solution that can be attained by the centralized schemes. However, we expect that in generic topologies the performance of the distributed rate allocation algorithms will be inferior to that of the centralized ones, as the myopic optimization performed by the network users does not always detect all the opportunities for inter-session packet combinations. 

\begin{figure*}[t!]
	\begin{center}
		\subfloat[] {\label{fig:distr_a}\includegraphics[width=0.49\textwidth]{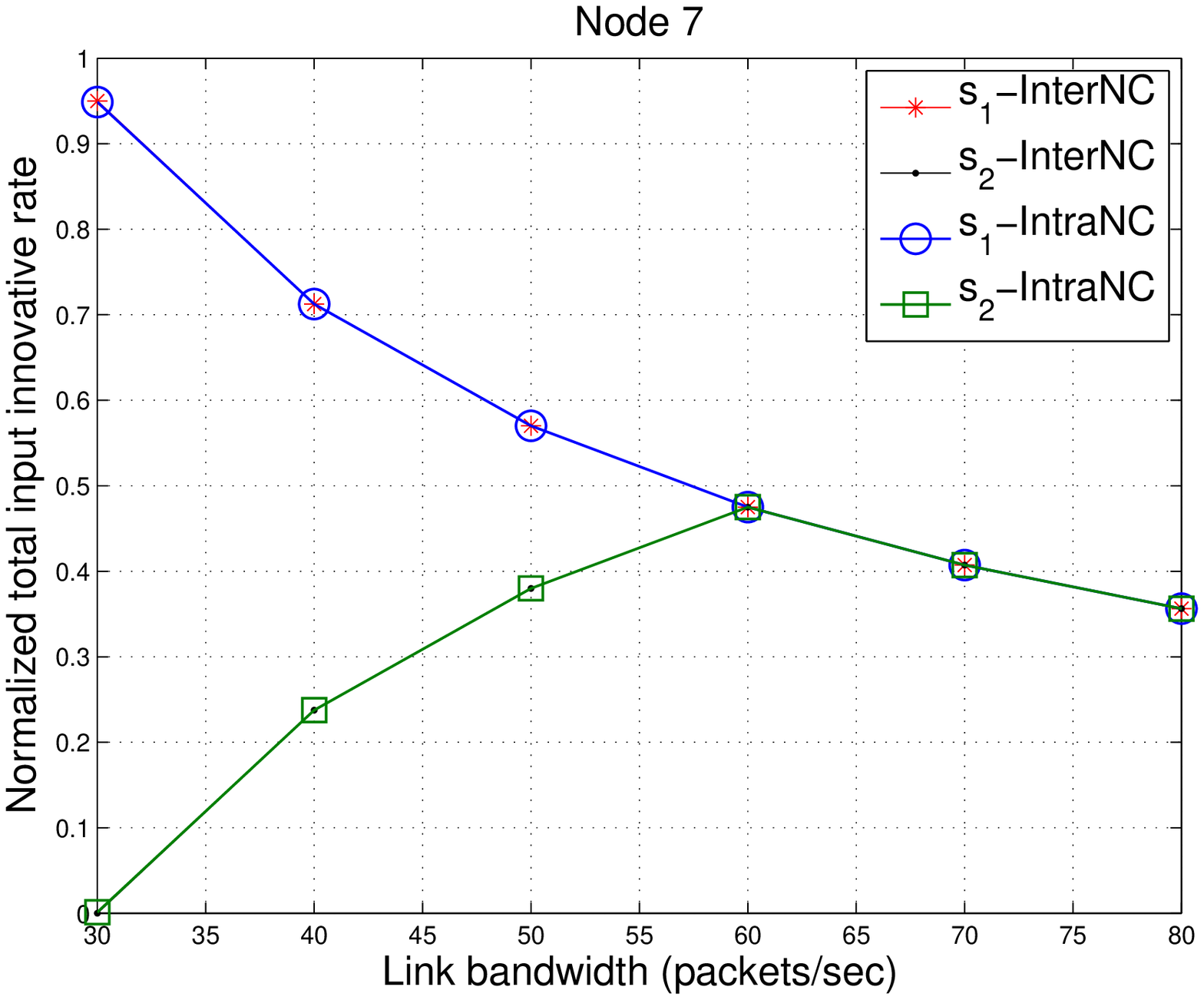}}  
		\subfloat[] {\label{fig:distr_b}\includegraphics[width=0.49\textwidth]{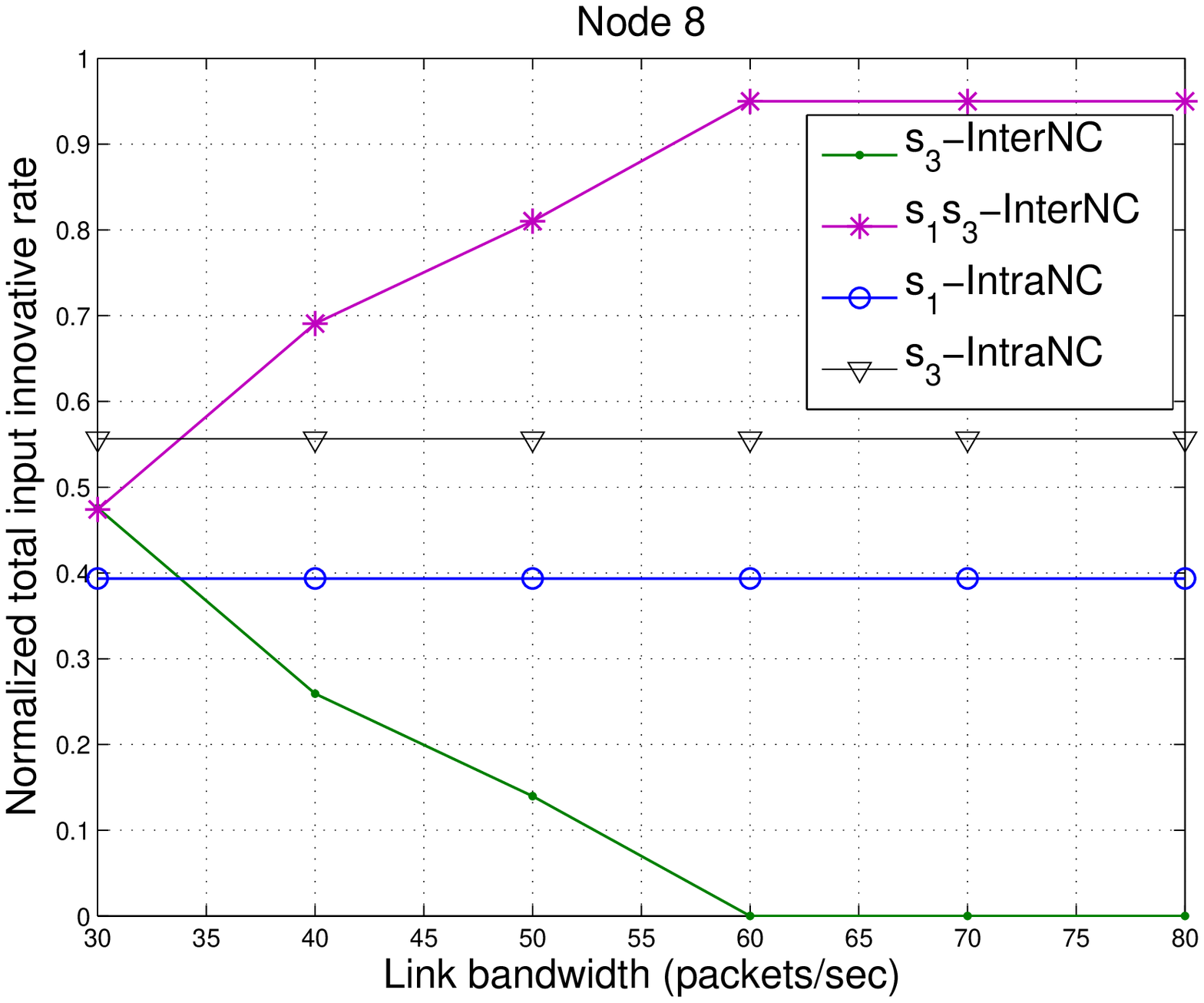}}  \\
		\subfloat[]{\label{fig:distr_c}\includegraphics[width=0.49\textwidth]{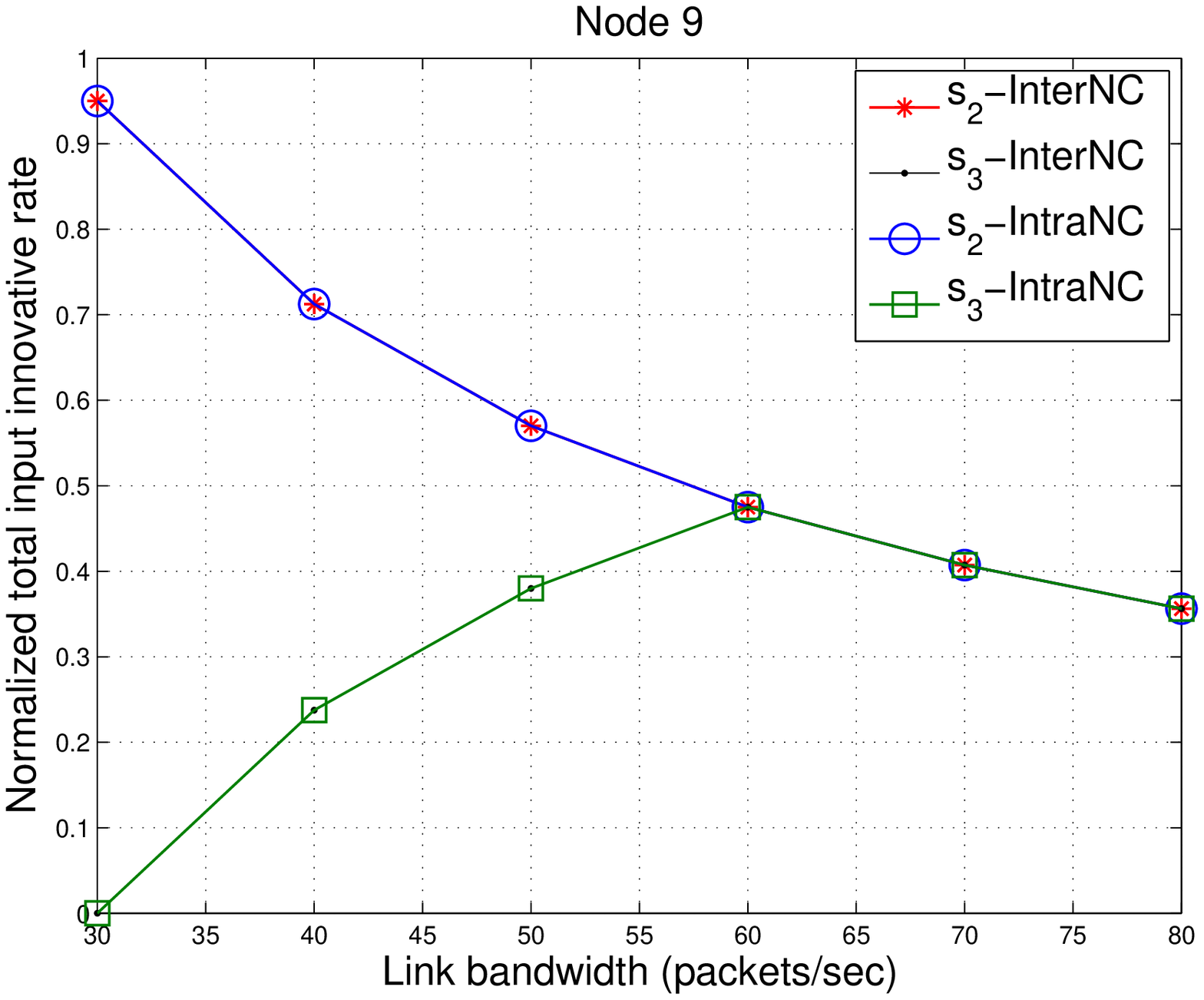}} 
	\end{center}
	\caption{Normalized total input innovative packet rate for nodes (a) $n_7$, (b) $n_8$, (c) $n_9$ versus the bandwidth of links connecting nodes $n_4, n_7$ and $n_6,n_9$ for the topology depicted in Fig.~\ref{fig:topology1}(a). The schemes under comparison are the distributed InterNC and the distributed IntraNC rate allocation algorithms. \label{fig:rate_distr_toynetwork}}
\end{figure*}

\begin{figure*}[t]
	\begin{center}
		\subfloat[] {\label{fig:centr_a}\includegraphics[width=0.49\textwidth]{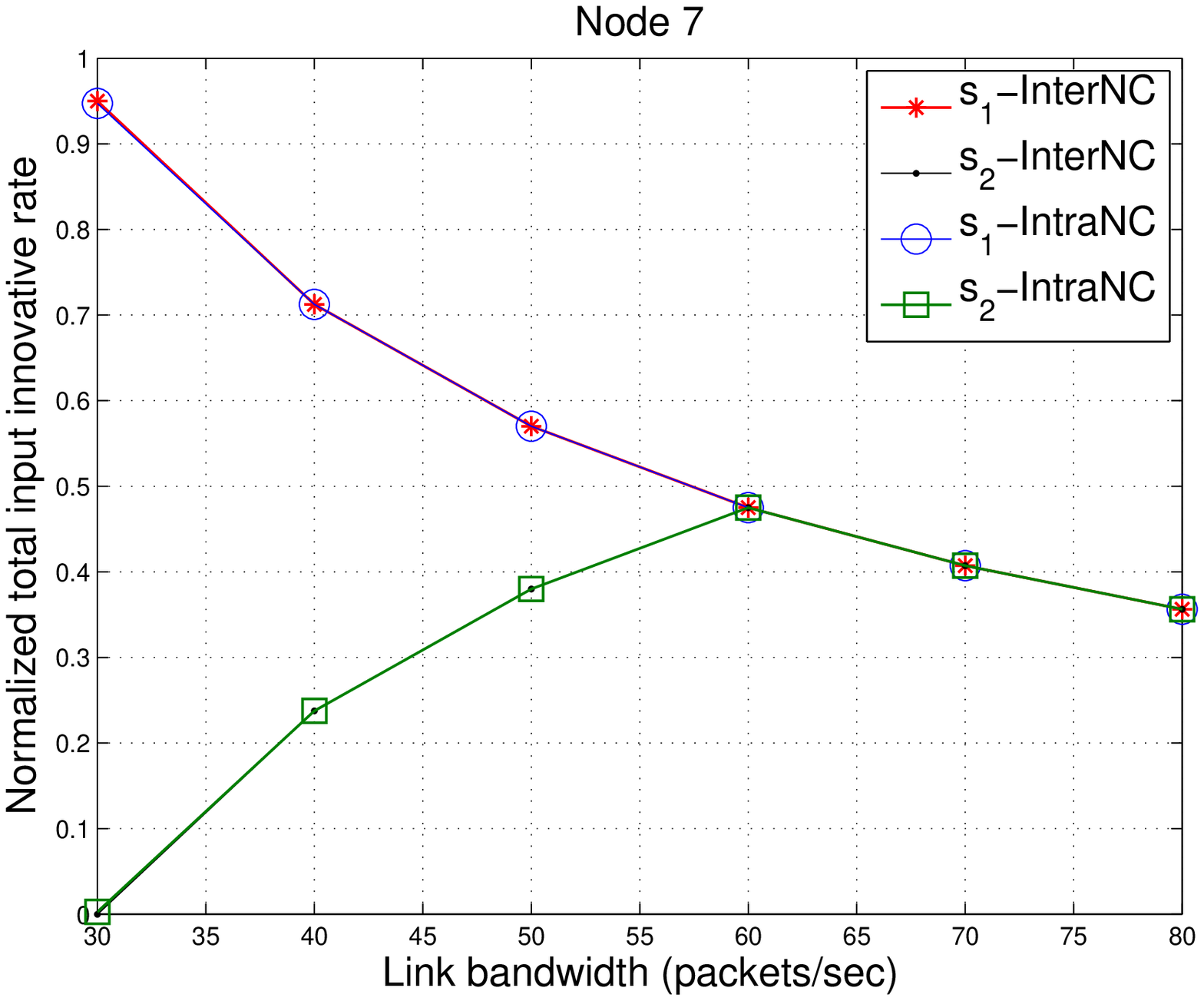}} 
		\subfloat[] {\label{fig:centr_b}\includegraphics[width=0.49\textwidth]{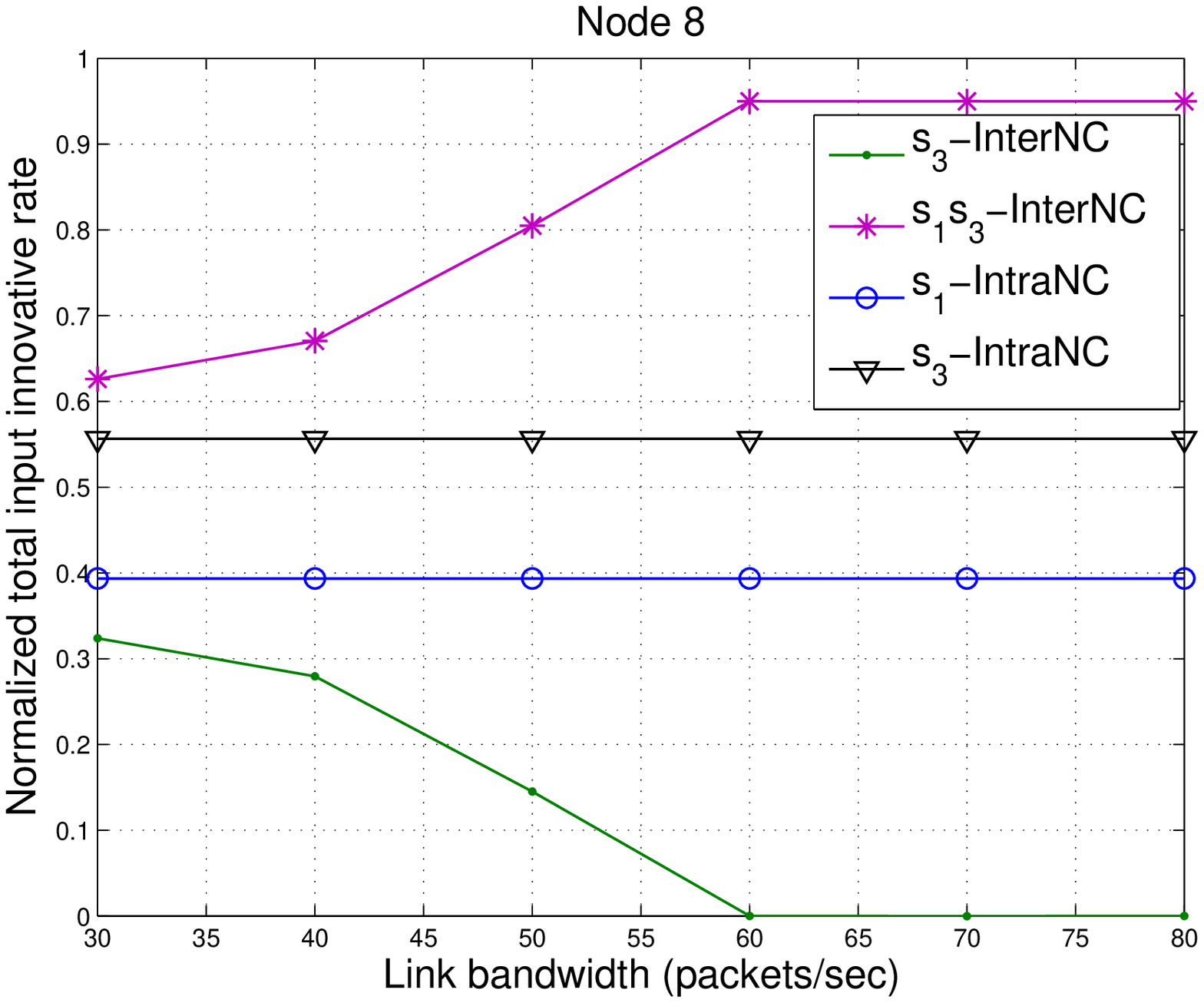}} \\
		\subfloat[]{\label{fig:centr_c}\includegraphics[width=0.49\textwidth]{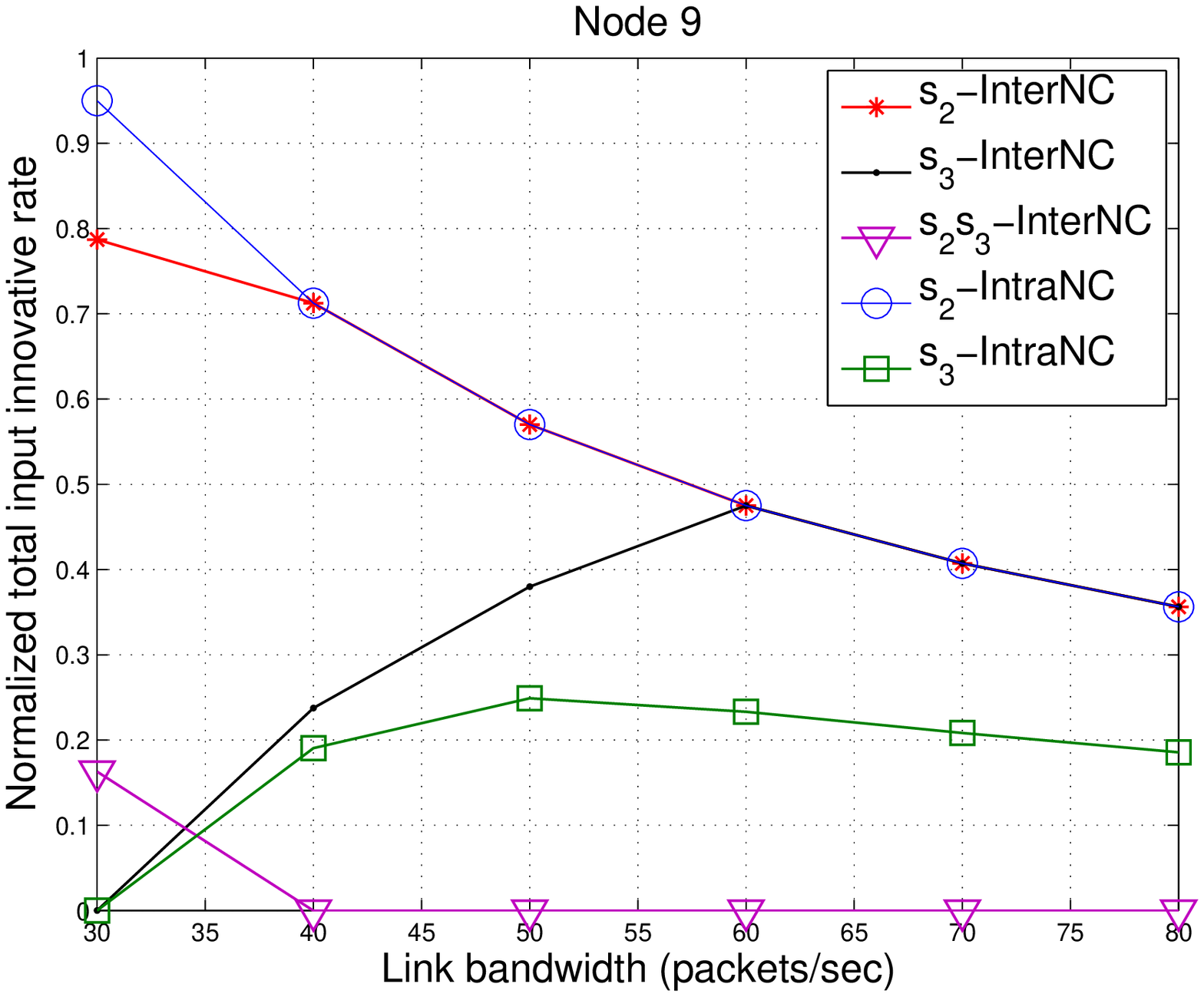}} 
	\end{center}
	\caption{Normalized total input innovative packet rate for nodes (a) $n_7$, (b) $n_8$, (c) $n_9$ versus the bandwidth of links connecting nodes $n_4, n_7$ and $n_6,n_9$ for the topology depicted in Fig.~\ref{fig:topology1}(a). The schemes under comparison are the centralized InterNC and the centralized IntraNC rate allocation algorithms. \label{fig:rate_centr_toynetwork} }
\end{figure*}

Our conclusions regarding the average decoding delay can be further supported by examining the innovative rate that is achieved by the schemes under comparison. Figs.~\ref{fig:rate_distr_toynetwork} and \ref{fig:rate_centr_toynetwork} illustrate the normalized total innovative input packet rate of nodes $n_7$, $n_8$ and $n_9$ for the distributed and centralized algorithms, respectively. The normalization is done with respect to the total input bandwidth of the user. In the figures, $s_j$ denotes a flow of intra-session network coded packets of session $s_j$, whereas $s_is_j$ represents the combined flow of inter-session network coded packets from sessions $s_i$ and $s_j$. The flows that are zero in the whole range of link bandwidths are omitted from the figures.

As we can notice from Figs.~\ref{fig:rate_distr_toynetwork}(b) and \ref{fig:rate_centr_toynetwork}(b), the link between nodes $n_5$ and $n_8$ has to be shared by the flows $s_1$ and $s_3$ when only intra-session network coding is allowed, as this is the only path from where nodes $n_{10}$ and $n_{12}$ can receive their requested flows. Thus, when the bandwidth of the links between nodes $n_4,n_7$ and $n_6,n_9$ increases, the average decoding delay of nodes $n_{10}$ and $n_{12}$ cannot be improved as they receive intra-session network coded packets at constant rates regardless of the bandwidth variations. The only reason for the slight improvement of the average delay that we observe in Fig.~\ref{fig:topology1}(b) is the additional supply of packets of session $s_2$ to node $n_{11}$ from node $n_7$, as can be seen by observing the rate curves in Figs.~\ref{fig:rate_distr_toynetwork}(a) and \ref{fig:rate_centr_toynetwork}(a).

When inter-session network coding is allowed, the average performance of the network is enhanced mainly by the combination of flows $s_1$ and $s_3$ on the bottleneck link between nodes $n_5$ and $n_6$. As we can see in Figs.~\ref{fig:rate_distr_toynetwork}(b) and \ref{fig:rate_centr_toynetwork}(b), the node $n_8$ allocates part of the input bandwidth to the combined flow $s_1s_3$ whereas the rest is allocated to the intra-session network coded flow $s_3$. As the node $n_9$ starts to provide more intra-session network coded packets of flow $s_3$ to node $n_{12}$ when the bandwidth increases, the percentage of rate for the combined flow on the bottleneck link increases and eventually the node $n_8$ requests only combined packets. At this point, both nodes $n_{10}$ and $n_{12}$ manage to receive their requested flows at the rate of the bottleneck link since they receive at the same rate the other component packets of the combined flow from nodes $n_7$ and $n_9$ respectively, and they are able to decode faster the session of their interest. Thus, we can see that the limitations imposed by the bottleneck link can be overcome by deploying inter-session network coding and utilizing the additional resources of the nodes for receiving packets that can help in decoding the combined sessions. 

It is worth noting that the rate allocation achieved by the distributed inter-session network coding algorithm is not identical to the one achieved by the centralized scheme for link bandwidth equal to 30 packets/sec, as can be seen in Figs.~\ref{fig:rate_distr_toynetwork}(b),  \ref{fig:rate_distr_toynetwork}(c),  \ref{fig:rate_centr_toynetwork}(b) and  \ref{fig:rate_centr_toynetwork}(c). This is attributed to the fact that the centralized algorithm has the full knowledge of the network topology. It can detect more opportunities for combining packets from different sessions, whereas the distributed scheme can only take advantage of the local network conditions. Finally, we can observe that for all schemes, the innovative rates and the average delay saturate as links' bandwidth reaches the value of 60 packets/sec. This is essentially the point where the system has reached the state where no other improvement can be achieved with either of the schemes.

Note that, since nodes $n_4$, $n_5$ and $n_6$ receive all their packets directly from the sources, they do not affect the average observed delay. The behavior of nodes $n_{10}$, $n_{11}$ and $n_{12}$ depends also on the rates available at nodes $n_7$, $n_8$ and $n_9$ as by the construction of the network they have sufficient download bandwidth in order to download all the packets that are available in the aforementioned nodes.

 \begin{figure*}[t!]
	\begin{center}
		\subfloat[]{\label{fig:topology2}\includegraphics[width=0.35\textwidth]{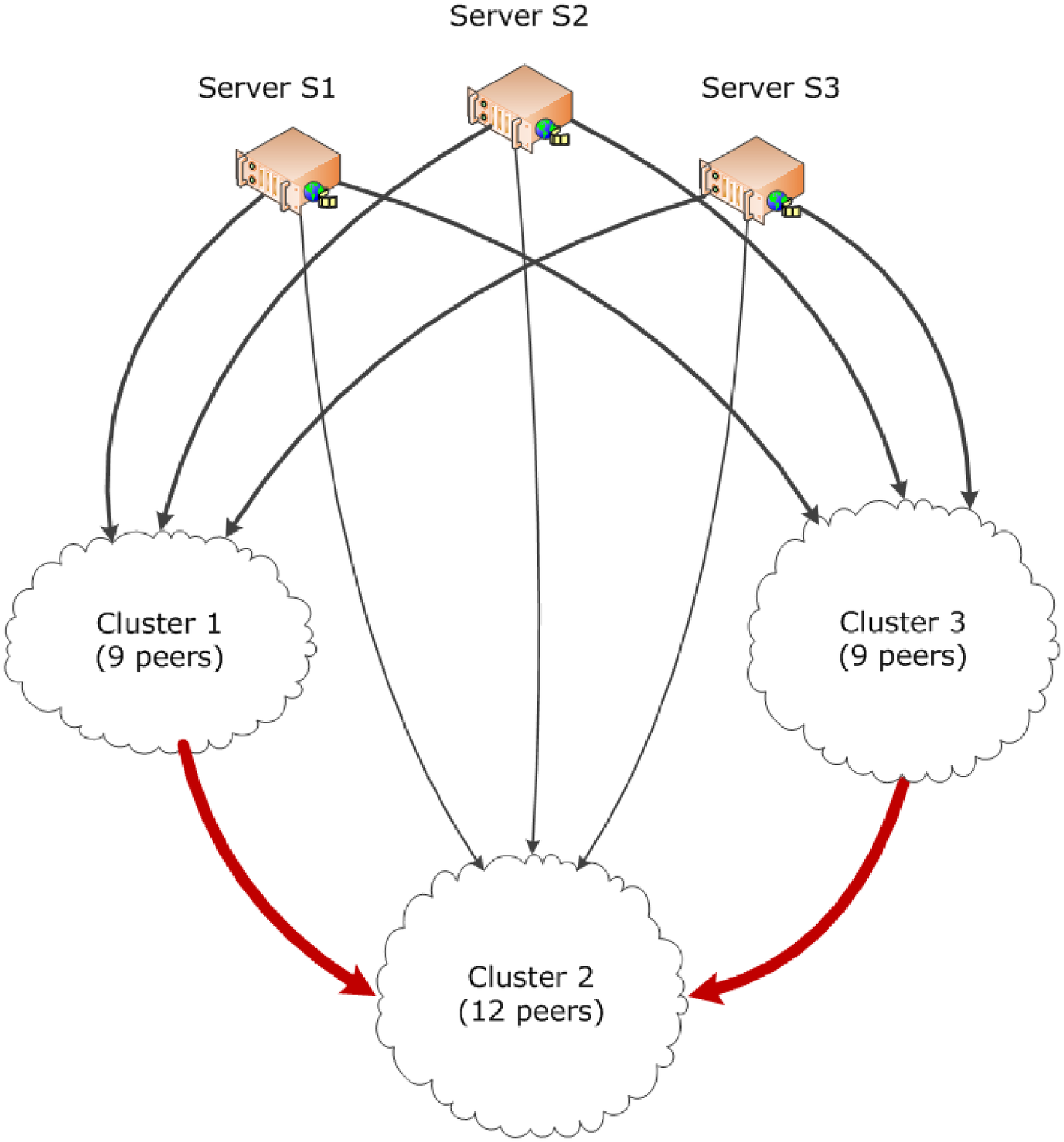}} \\
		\subfloat[]{\label{fig:delay_clust}\includegraphics[width=0.49\textwidth]{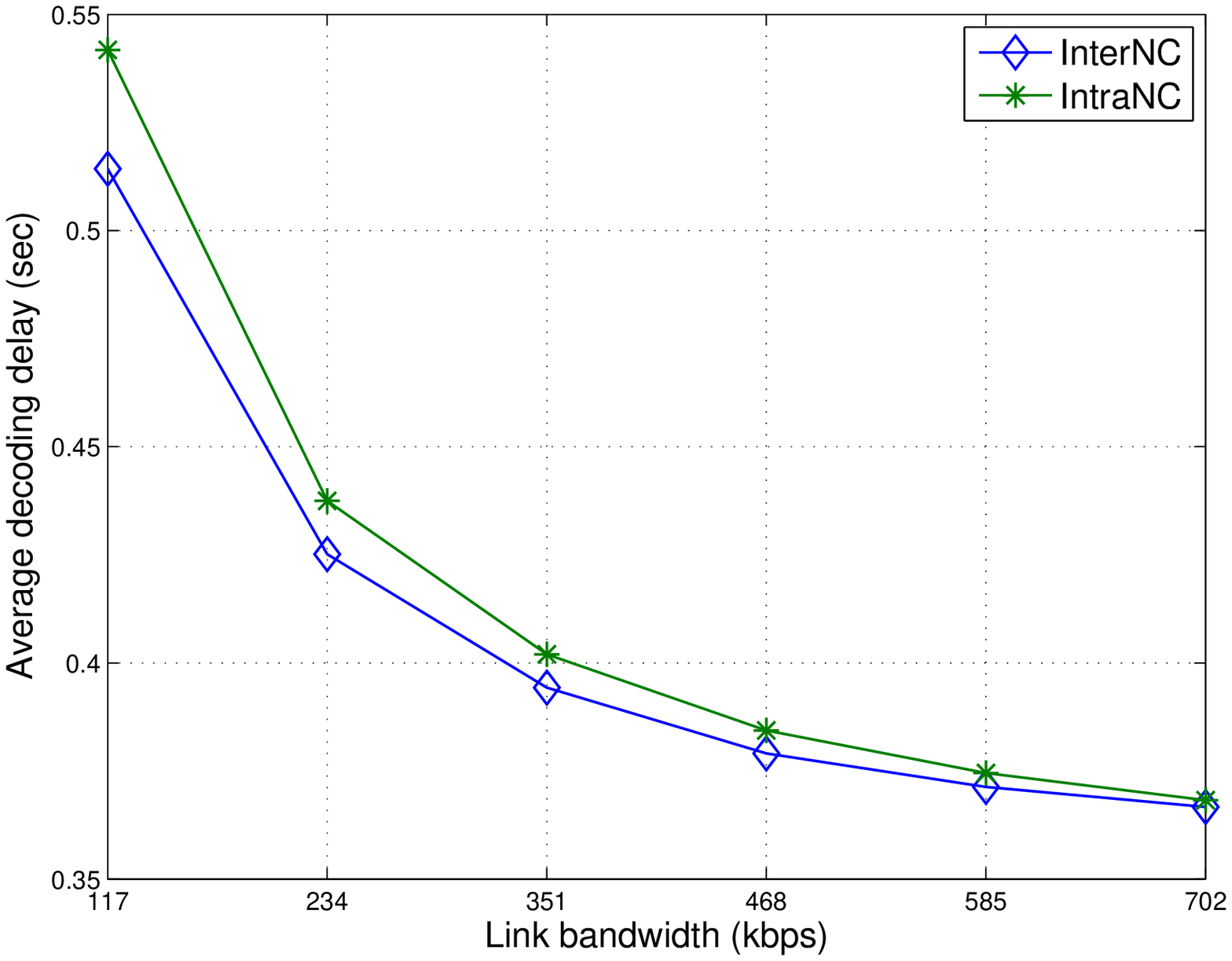}}
		\subfloat[]{\label{fig:delaypercluster}\includegraphics[width=0.49\textwidth]{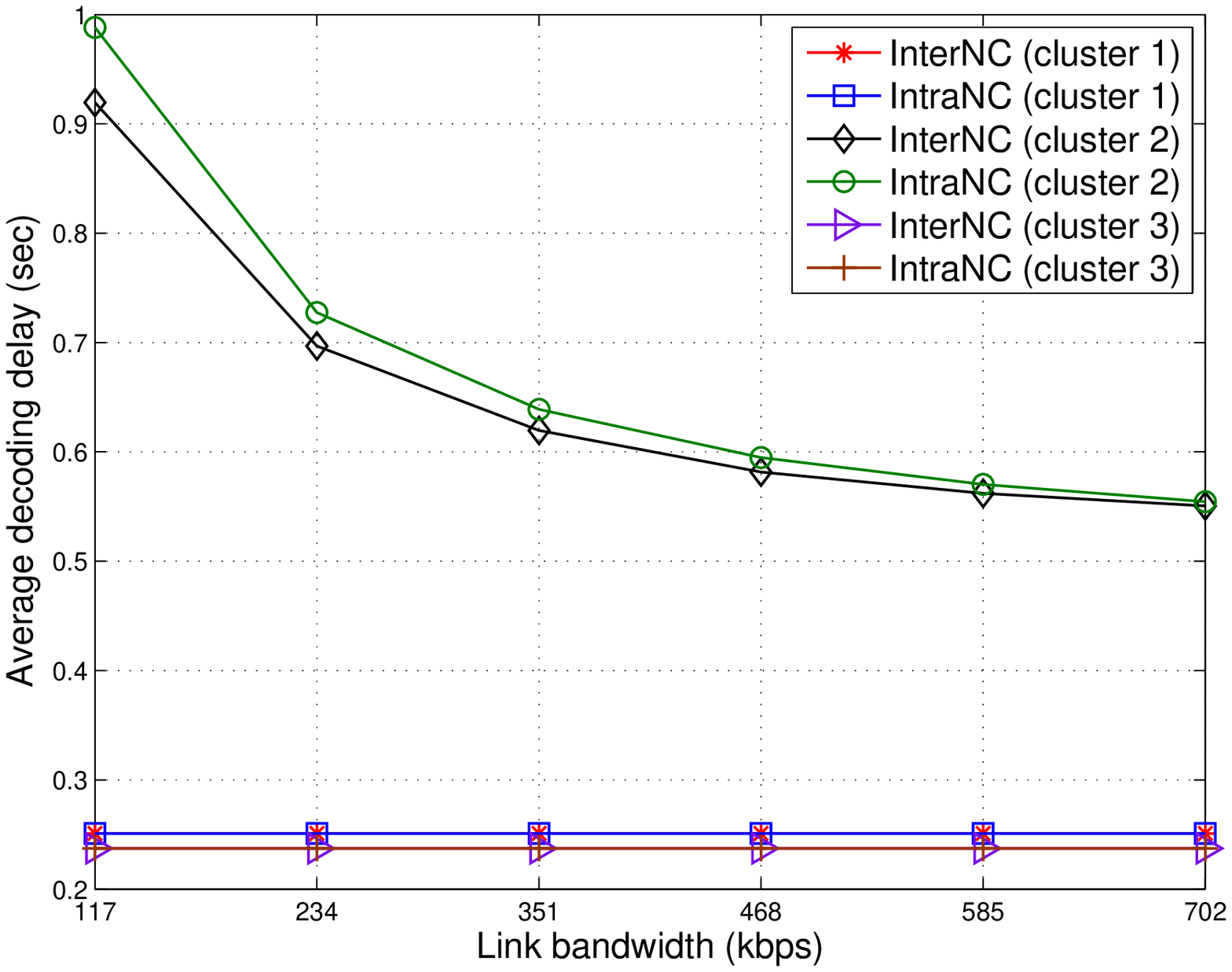}} 
	\end{center}
	\caption{Performance comparison of the InterNC and IntraNC algorithms with respect to the average decoding delay as a function of the links' bandwidth in a cluster network. (a) Cluster network topology, (b) average decoding delay for the whole network and (c) average decoding delay for each cluster of the network separately.\label{fig:cluster}}	
\end{figure*}


 \subsection{Clustered networks}
 \label{sec:clusters}

In this set of experiments, we evaluate the performance of the proposed scheme for the clustered network depicted in Fig.~\ref{fig:cluster}(a). This network consists of three server nodes and 30 client nodes. The clients are organized in 3 clusters of 9, 12 and 9 nodes respectively. Each cluster is an irregular directed network generated from a regular network by removing and shifting randomly some of the links \cite{ThomosTCSVT2010}. The pruning and shifting probabilities are set to 40\% and 20\% respectively.  Every user is assigned one of the data sources. The selection of the sources is done uniformly at random. The clusters 1 and 3 are connected directly to the servers with links that have a capacity of 468 kbps each, whereas the cluster 2 is connected to the clusters 1 and 3 through links with a capacity that varies in the interval $[117, 702]$ kbps. Moreover, the cluster 2 receives some packets directly from the sources through low speed links that have capacity of 468 kbps.  Finally, the nodes within all the clusters are interconnected with high speed links of 1.6 Mbps. The packets size is fixed to 1500 bytes including the network coding header. Again we consider that the block sizes for all data sources are equal to 10 packets. All the results in this section are averages of 10 random realizations of the network.

Fig.~\ref{fig:cluster}(b) illustrates the average decoding delay for the clustered network depicted in Fig.~\ref{fig:cluster}(a) with respect to the bandwidth of the links that connect cluster 2 to clusters 1 and 3. The schemes under comparison is the proposed distributed InterNC rate allocation algorithm and the baseline distributed IntraNC scheme. We can observe that, by allowing nodes to combine data from different sessions, we can achieve lower decoding delay times than those that can be achieved with intra-session network coding only. As presented in Fig.~\ref{fig:cluster}(c), the gain is observed in cluster 2 that does not have sufficient resources to provide intra-session network coded packet to all the users, contrarily to clusters 1 and 3 where all the users are able to acquire all the packets directly from the sources. Thus, inter-session network coded packets are requested on the bottleneck links connecting cluster 2 to clusters 1 and 3 in order to serve more users in the network, whereas the additional packets that are provided through the low capacity links that connect cluster 2 to the sources are used to decode faster the combined packets.

 
 \section{Video streaming simulations}
\label{sec:video}

In this section, we analyze the performance of the proposed rate allocation algorithm in video streaming simulations. The packets of a video sequence are typically grouped into several blocks of packets with similar decoding deadlines, {\em i.e., generations} \cite{ChouPractical03}, and the intra- or inter-session network coding operations are performed on packets that belong to the same generation. This is due to the fact that in network coded systems, the packets that belong to the same generation are decoded simultaneously. Therefore, the generation has to be decoded before the most urgent packet of the generation expires.

The presence of multiple temporally consecutive generations necessitates scheduling mechanisms that are responsible for the timely delivery of the generations to the users. Thus, we first propose a scheduling mechanism that regulates the transmission of multiple generations in combination with the optimal rate allocation strategy described above. We then evaluate the proposed framework in different video delivery scenarios using the network simulator NS-3 \cite{ns3}.


\subsection{Multiple generations scheduling}
\label{sec:multigeneration}
 
We consider the system setup described in Section \ref{sec:systemoverview}. The source packets transmitted by the source $s$ are grouped into generations of size $N_{s}$. The $i$-th generation is identified by the generation index $G_{i}$ and has a decoding deadline denoted as $T_{i}$. In order to coordinate the transmission of multiple generations, every node keeps track of the generation that has to be transmitted on each outgoing link, and forwards packets of this generation at rates determined by the rate allocation algorithm presented in Section \ref{sec:ratealloc}. 

The generation indexes on the outgoing links are updated according to a schedule which is decided based on the feedback provided by the children nodes. Let us focus on one of the network nodes and let us denote as $\tau_{i}$ the time instant when the node sends a request to its parents to update the generation index on the node's input links to $G_{i}$. At time $\tau_{i}$, along with the request for the generation index update, the node schedules the next request to be transmitted at time $\tau_{i+1}$, when the generation index on its input links will be updated to $G_{i+1}$. Initially, the time $\tau_{i+1}$ is set equal to the decoding deadline $T_{i}$ of the generation $G_{i}$. However, the request can be rescheduled to an earlier time instant $\tau_{i+1}^{\prime} < \tau_{i+1}$ as soon as the following two conditions are fulfilled: i) the node has received a feedback message from all the children nodes indicating that they have either decoded or decided to skip generation $G_{i}$ and ii) the node has either decoded or has decided to skip generation $G_{i}$. In that case, the node immediately requests its parents to update the generation index to $G_{i+1}$ on all its input links. Otherwise, if the two above conditions are not fulfilled before the time instant $\tau_{i+1}$, the request for the next update of the generation index is sent according to the original schedule. In both cases, the node schedules the next request for an update of the generation index at time $\tau_{i+2} = T_{i+1}$, where $T_{i+1}$ is the decoding deadline of the generation $G_{i+1}$. Note that the generation $G_{i}$ may not become available immediately when the node requests its parents to update the generation index on its input links to $G_{i}$, as the parent nodes may still be requesting earlier generations.

The decision to skip a generation is taken at every node independently based on the estimation of the average time that is required to receive and decode one generation of packets. In order to decide at time $\tau_{i}$ whether to skip the generation $G_{i}$, every node first updates its estimation of the average decoding time. This update is performed by recursively updating the sample mean of the approximate decoding times of previously transmitted generations. Let us denote as $\delta t_{i-1}$ the approximate decoding time of generation $G_{i-1}$. If the generation $G_{i-1}$ was decoded, the decoding time $\delta t_{i-1}$ is calculated as the difference between the time instant $\tau_{i-1}^{d}$ when the generation was decoded and the time $\tau_{i-1}^{f}$ when the first packet of generation $G_{i-1}$ was received, since in general the generation $G_{i-1}$ does not become available to the node immediately after it has been requested.  If the generation $G_{i-1}$ was not decoded before the time instant $\tau_{i}$, the approximate decoding time $\delta t_{i-1}$ is set equal to $\alpha (\tau_{i}-\tau_{i-1}^{f})$, where $\alpha > 1$. The multiplicative term $\alpha$ compensates for the fact that the time elapsed between two consecutive generation update events was not sufficient for the node to decode the transmitted generation.\footnote{The value of $\alpha$ is determined by experimentation.} The new sample $\delta t_{i-1}$ is then used to update the node's estimation of the average decoding time. Note that the node updates its estimation of the average decoding time only if it has not decided to skip generation $G_{i-1}$. Once the estimation of the average decoding time has been updated, the node compares the average decoding time to the time interval $\tau_{i+1} - \tau_{i}$, which represents the maximum available time for decoding generation $G_{i}$. If the average decoding time is larger than this time interval, the node makes the decision to skip generation $G_{i}$ and sends a feedback message to its parents informing about this decision. The skipping policy permits the node to skip a generation and to save resources in order to decode subsequent generations.


\subsection{Simulation results}
\label{sec:ns3simulations}

We now evaluate our distributed rate allocation algorithm, combined with the scheduling scheme proposed in Section \ref{sec:multigeneration}, for the transmission of video sequences. For the evaluation, we encode the {\em Carphone}, {\em Foreman} and {\em Container} QCIF format video sequences with the H.264/AVC video compression standard \cite{H.264Overview} at rate 240kbps. Each sequence consists of 300 frames that are repeated in order to obtain sequences of 40 sec duration. The frames are encoded as {\em IPPPP...} with a frame rate set to 30fps. The size of the GOP is 30 frames and the average PSNR per frame is 39.14 dB, 38.8 dB and 42.85 dB for the three sequences, respectively. Each generation consists of 20 packets and corresponds to a GOP. The payload of each packet is 1500 bytes. Each packet is augmented with a header of 81 bytes that contains packet information, {\em i.e.}, network coding coefficients, packet type, generation number and time stamp. The proposed framework is simulated with the help of the network simulator NS-3 \cite{ns3}. All the results are averages of 20 simulations.

We first evaluate the proposed framework for the network topology depicted in Fig.~\ref{fig:topology1}(a). The bandwidth of the links that originate from the sources, as well as of the link connecting nodes $n_{5}$ and $n_{8}$ is set to 607 kbps. The bandwidth of the links that originate from nodes $n_{7}$, $n_{8}$ and $n_{9}$ is set to 1214 kbps. The packet loss rate is set to 5\% on all links. 

\begin{figure*}[t!]
	\begin{center}
		\subfloat[]{\label{fig:AvgDecGenVsBwd}\includegraphics[width=0.49\textwidth]{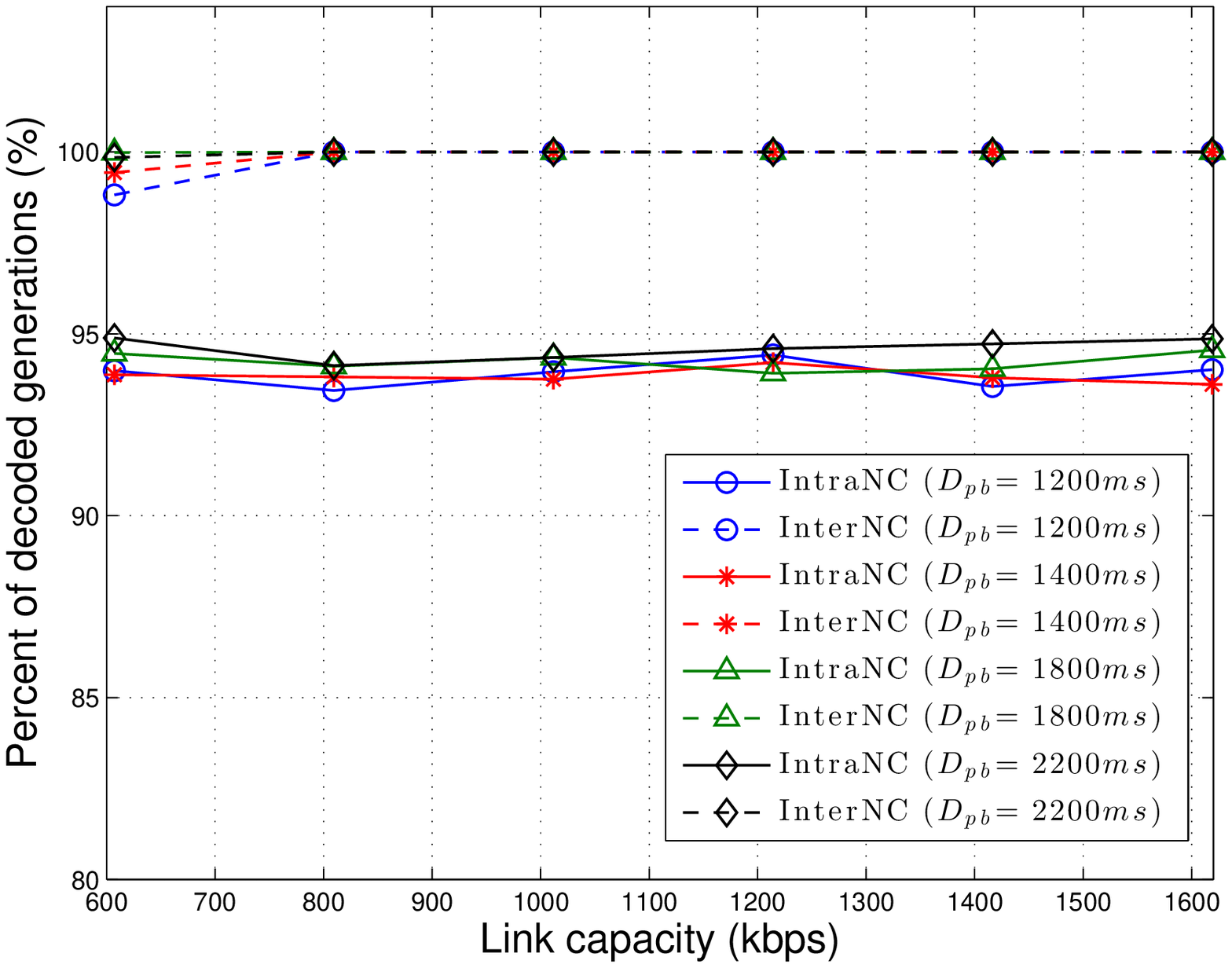}} 
		\subfloat[]{\label{fig:AvgPsnr}\includegraphics[width=0.49\textwidth]{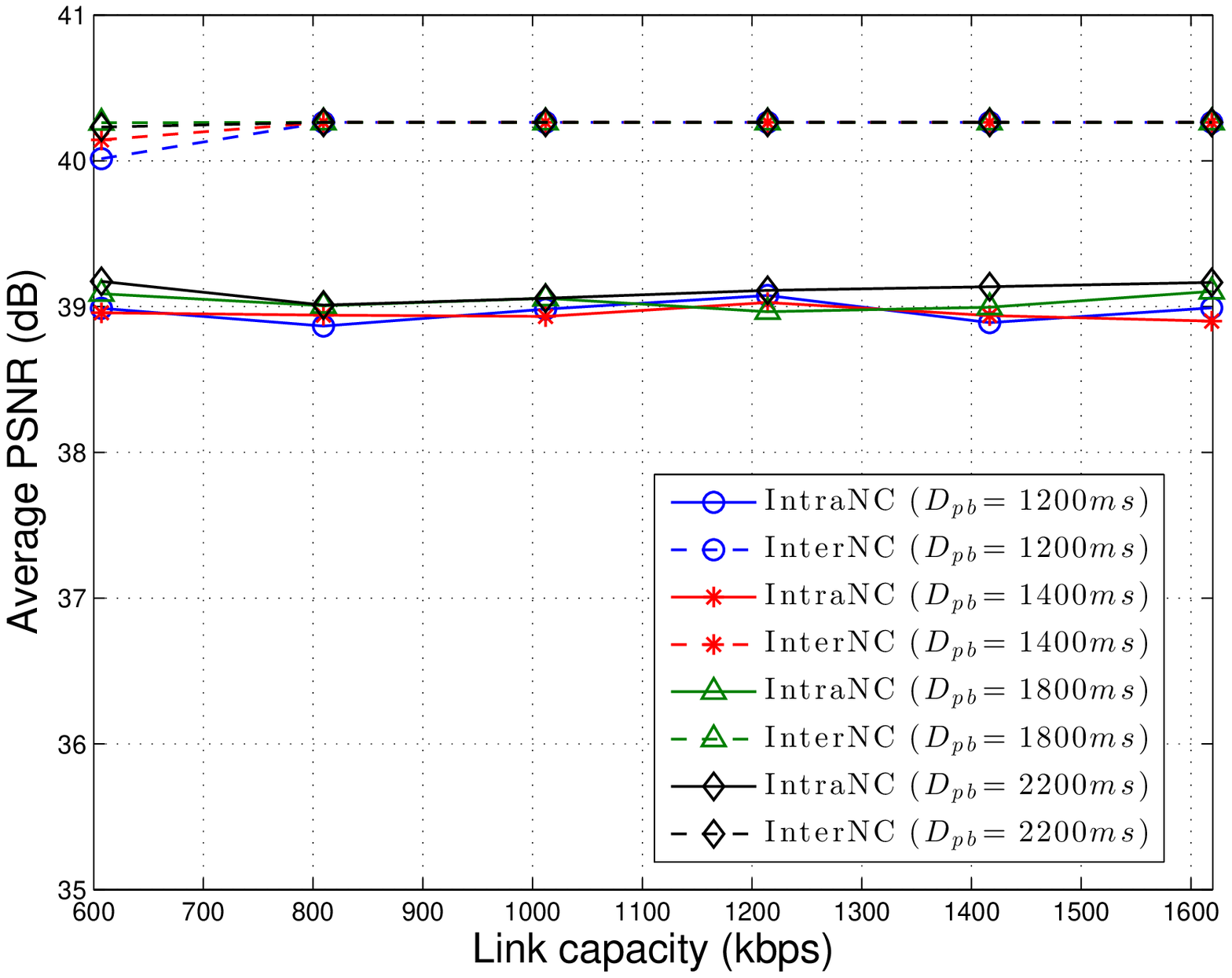}} \\ 
		\subfloat[]{\label{fig:AvgPsnrVsTimeN12PbD1200}\includegraphics[width=0.49\textwidth]{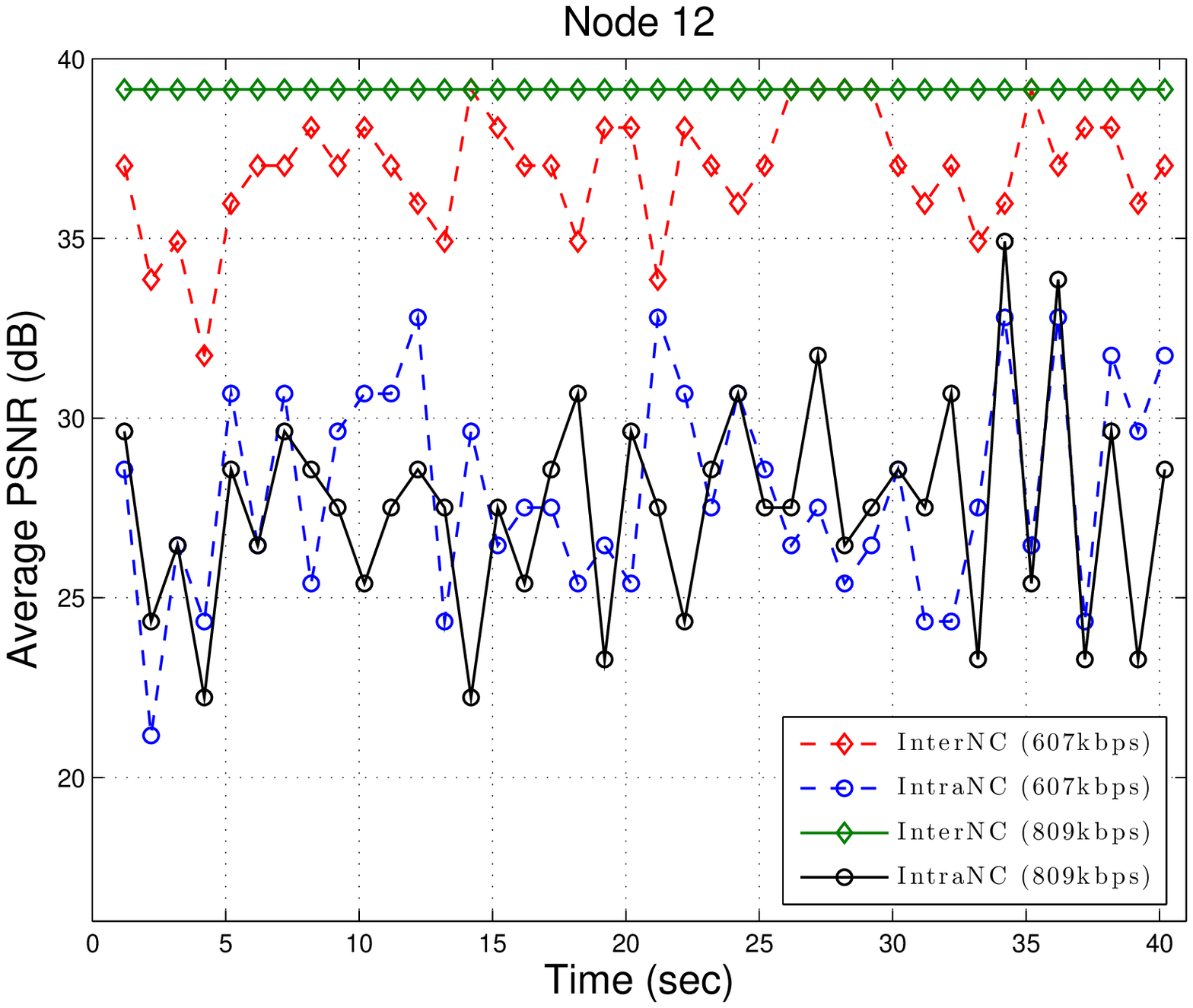}}
	\end{center}
	\caption{(a) Average percent of decoded generations and (b) average PSNR of the Y-component of the transmitted video sequences after decoding for different values of playback delay $D_{pb}$ as a function of links' bandwidth for the network depicted in Fig.~\ref{fig:topology1}(a). (c) Evolution of the average PSNR with time at node $n_{12}$ of the network depicted in Fig.~\ref{fig:topology1}(a) for playback delay $D_{pb} = 1200\mbox{ms}$ and  two different values of  links' bandwidth. The node $n_{12}$ requests the {\em Carphone} video sequence with average PSNR per frame equal to 39.14 dB. \label{fig:AvgDecGenAndPsnr}}
\end{figure*}

Fig.~\ref{fig:AvgDecGenAndPsnr}(a) illustrates the percent of decoded generations averaged over the number of nodes in the network versus the bandwidth of the links connecting nodes $n_{4}$, $n_{7}$ and $n_{6}$, $n_{9}$ for different values of playback delay $D_{pb}$. The playback delay is defined as the time allowed for initial data buffering before the start of the playback. We can observe that the distributed InterNC rate allocation scheme achieves better performance in terms of the average number of decoded generations compared to the IntraNC scheme as it provides lower decoding delays, thus enabling the decoding of generations prior to their expiration deadlines. For high values of links'  bandwidth, the nodes are able to decode the full video sequence with the inter-session network coding based rate allocation scheme, while for low bandwidth values the nodes decode on average more than 95\% of the video sequences. On the contrary, the decoding delay achieved with the IntraNC rate allocation scheme is not sufficient in order to guarantee a smooth playback of the video sequences. Nodes with high decoding delays are forced to skip a significant number of generations in order to be able to decode at least a part of the video sequence that they request. The decoding delays of these nodes are not affected by the increase in the links' bandwidth as the IntraNC rate allocation scheme cannot take advantage of the additional network resources. Thus, the performance of the intra-session network coding based scheme remains invariant with the increase in the links' bandwidth. We can also see that the value of the initial playback delay does not influence significantly the average performance of the network. Larger values of playback delay permit nodes with limited resources to decode more generations in the beginning of the transmission process, which improves slightly the overall performance. However, even higher values of playback delay are not sufficient to enable timely delivery of subsequent generations for nodes with scarce resources. 

In Fig.~\ref{fig:AvgDecGenAndPsnr}(b), we present the average PSNR of the Y-component of the transmitted video sequences after decoding at the nodes, as a function of the links' bandwidth for different values of playback delay. We set the average PSNR of the generations that could not be decoded to 18 dB for the {\em Carphone} and {\em Foreman} sequences and to 19 dB  for the {\em Container} sequence. The results show that, with the InterNC rate allocation scheme, the network users display the video at better quality than with the IntraNC rate allocation scheme; in the latter case, the decoding delays are too high to guarantee a constant quality playback for all the nodes. As an example, we illustrate in Fig.~\ref{fig:AvgDecGenAndPsnr}(c) the average video quality obtained at node  $n_{12}$ versus time for a playback delay $D_{pb} = 1200\mbox{ms}$ and different values of the links' bandwidth. Each point in the curves is obtained by averaging the corresponding PSNR values in 20 simulations. The node $n_{12}$ is the node that is most affected by the bottleneck link between nodes $n_{5}$ and $n_{8}$. We can see that the IntraNC rate allocation scheme performs poorly and does not improve as the links' bandwidth increases. The average video quality presents significant fluctuations over time. In contrast, the average video quality obtained with the InterNC rate allocation scheme remains more stable over time and improves significantly as the links' bandwidth increases.

We now further evaluate the proposed framework for one random realization of the clustered network topology depicted in Fig.~\ref{fig:cluster}(a). The links that connect the servers to the clusters have capacity of 759 kbps, whereas the cluster 2 is connected to the clusters 1 and 3 through links with a capacity that varies in the interval [190,\;1138] kbps. The users within each cluster are interconnected with high speed links of 2.6 Mbps. The packet loss rate is set to 5\%.

\begin{figure}[t]
\begin{center}
	\includegraphics[width=0.49\textwidth]{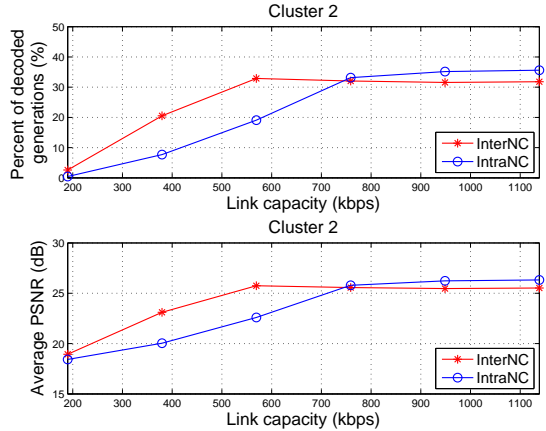}
	\caption{Average percent of decoded generations (top) and average PSNR of the Y-component of the transmitted video sequences after decoding (bottom) at the network nodes for the second cluster of the network topology depicted in Fig.~\ref{fig:cluster}(a) as a function of the links' bandwidth for playback delay $D_{pb} = 1400\mbox{ms}$.  \label{fig:AvgDecGenVsBwdPbD1400clust}}
\end{center}
\end{figure}

Fig.~\ref{fig:AvgDecGenVsBwdPbD1400clust} (top) depicts the average percent of generations decoded by the network nodes that belong to the cluster 2 as a function of the bandwidth of the links that connect the cluster 2 to the clusters 1 and 3. The playback delay is set to $D_{pb} = 1400$ms. We can observe that the performance of both the IntraNC and the InterNC rate allocation schemes improves as the links' bandwidth increases. However, the performance of the InterNC rate allocation scheme stays superior to the performance of the IntraNC rate allocation scheme for low values of the links bandwidth. This is due to the more efficient exploitation of the additional resources provided by the links that directly connect some of the nodes in the cluster 2 to the sources, as we have discussed in Section \ref{sec:clusters}. The performance of the two schemes is similar for higher values of bandwidth, where the performance of certain nodes with scarce resources cannot be further improved even with inter-session network coding. Similar conclusions can be reached by observing the average PSNR of the video sequences after decoding at the nodes of the cluster 2, which is illustrated in Fig.~\ref{fig:AvgDecGenVsBwdPbD1400clust} (bottom).

Finally, we would like to remark that the initial playback delay does not influence significantly the performance of the proposed schemes. We have repeated the simulations for playback delay values equal to 1200ms, 1400ms, 1800ms and 2200ms. Though higher values of the initial playback delay permit users to decode more generations in the beginning of the transmission process, nodes with scarce resources are not able to decode the subsequent generations even for large values of playback delay. Furthermore, we have omitted the results for clusters 1 and 3, since these clusters have sufficient resources to obtain the video sequences at optimal quality.


\section{Conclusions}
\label{sec:conclusions}

We have proposed a novel distributed rate allocation algorithm for delivery of multiple concurrent sessions in wireline mesh networks. The algorithm is based on inter-session network coding. The network users decide locally on the optimal coding decisions and rates for each combination of packets that they request from their parents. The decisions are based on the minimization of the average decoding delay of the node and its children nodes and require only a minimal communication overhead. We show that the initial non-convex rate allocation problem can be decomposed into a set of simpler convex problems with the help of a new equivalent flow representation. The final rate allocation can then be obtained by combining the results of each of the subproblems. The evaluation of the proposed algorithm demonstrates the benefits of utilizing inter-session network coding in terms of the decoding delays and efficient exploitation of network resources. Simulation results show that the proposed algorithm is capable of eliminating the bottlenecks and reducing the decoding delay of users with limited resources. In the context of video transmission, it enables the timely delivery of video data to the network users, hence leads to better average video quality.

\bibliographystyle{IEEEtran}
\bibliography{netcoding}

\end{document}